\documentclass[aps,prd,onecolumn,groupedaddress,nofootinbib,amssymb,nopacs
]{revtex4}
\usepackage[dvips]{graphicx}
\usepackage{amssymb}
\usepackage{amsmath}
\usepackage{graphicx,color}
\usepackage{amsfonts}
\usepackage{bm}
\usepackage{cancel}
\usepackage{comment}

\allowdisplaybreaks[4]
 
\usepackage{hyperref}

\hypersetup{
linktocpage=true,
colorlinks=true,
linkcolor=blue,          
citecolor=mygreen,        
filecolor=blue,      
urlcolor=blue           
}  
\usepackage{color}
\definecolor{mygreen}{rgb}{0.0,0.75,0.0}

\usepackage{etoolbox}
\appto\UrlBreaks{\do\-}
\PassOptionsToPackage{hyphens}{url}
\usepackage[hyphenbreaks]{breakurl} 

\begin{document}

\tolerance=5000

\title{Reconstruction of the quintessence dark energy potential from a Gaussian process}

\author{
Emilio Elizalde$^{1}$\footnote{E-mail: elizalde@ice.csic.es}, 
Martiros Khurshudyan$^{1,2,3}$\footnote{E-mail: khurshudyan@ice.csic.es},
K. Myrzakulov$^{2,3}$\footnote{Email: krmyrzakulov@gmail.com}
and
S. Bekov$^{2,3}$\footnote{Email: ss.bekov@gmail.com}
}
\affiliation{
$^{1}$ Consejo Superior de Investigaciones Cient\'{\i}ficas, ICE/CSIC-IEEC,
Campus UAB, Carrer de Can Magrans s/n, 08193 Bellaterra (Barcelona) Spain \\
$^{2}$  Eurasian National University, Nur-Sultan 010008, Kazakhstan\\
$^{3}$ Ratbay Myrzakulov Eurasian International Centre for Theoretical Physics, Nur-Sultan 010009, Kazakhstan\\
}

\begin{abstract}
  
The quintessence dark energy potential is reconstructed in a model-independent way. Reconstruction relies on a Gaussian process and on available expansion-rate data. Specifically, 40-point values of $H(z)$ are used, consisting of a 30-point sample deduced from a differential age method and an additional 10-point sample obtained from the radial BAO method. Results are obtained for two kernel functions and for three different values of $H_{0}$. They shed light on the $H_{0}$ tension problem for a universe described with quintessence dark energy. They are also a clear indication that the tension has to do with the physical understanding of the issue, rather than  being just a numerical problem with statistics. Moreover, the model-independent reconstruction of the potential here obtained can serve as a reference to constraint available models and it can be also used as a reference frame to construct new ones. Various possibilities, including $V(\phi) \sim e^{-\lambda \phi}$, are compared with the reconstructions here obtained, which is notably the first  truly model independent reconstruction of the quintessence dark energy potential. This allows to select new models that can be interesting for cosmology. The method can be extended to reconstruct the potential of related dark energy models, to be considered in future work.

\end{abstract}


\maketitle

\section{Introduction}\label{sec:INT}

Modern cosmology is rich in interesting situations reflecting how our previous knowledge of the Universe need be modified to accommodate new observations. The $H_{0}$ tension problem is one of those, in that it points out a huge difference between the early time measurements (e.g., data from Cosmic Microwave Background(CMB) surveys and from Baryon acoustic oscillations(BAO)) and late time measurement (e.g., coming from Type Ia Supernovae (SNe Ia) and $\mathrm{H}(\mathrm{z}))$ for the Hubble constant $H_{0}$ \cite{H01}, \cite{H02}. Various interesting proposals on how the problem could be solved have appeared in the literature, as \cite{H0Start}-\cite{H0End} (and references therein). 

In the last several years, we have witnessed significant technological developments, which have helped us to improve the data collection process, analysis, and storage capacities by orders of magnitude. But without the possibility to do direct experiments with the Universe it is still difficult to deal with some problem. These shortcoming also appear in other research fields, having in common that it would  seem at first sight in all cases that we have, in principle, very advanced technological developments to tackle any issue thoroughly. However, why could we not make significant progress in solving some long-standing problems? Is it because of an issue with our understanding of what data mean? Is it a problem with our specific understanding of how to collect and correctly use these data? Eventually, is there a problem with the model-construction strategies to reflect our understanding of what the observational data say? Could this reflect that we cannot avoid a bias when we link a model with the data? Can these discrepancies be meaningful and  helpful? Imagine we have collected data and  want to  construct, directly from these data, a feasible model that produces them, but without relying on any preset theory.  Is this possible at all? 

In other words, is it possible to infer knowledge directly from the observed data, not referring to a specific model to compare them with? In a sense, some of the above  questions can be answered using Machine Learning. Even more, there is a solid belief, that eventually a huge part of the above mentioned questions can be answered with these techniques. But, what is exactly what machine learning aims at and why is it nowadays one of the top and more interesting research fields? Machine Learning (ML) tries to do the following: It does not start from questions but, on the contrary, having the answers it allows to find the questions (the models) explaining what we have in terms of experimental data. The implementation of various learning algorithms that have proven to be able to successfully tackle different problems, explains why recently the attention to ML has been increasing significantly. Moreover, long-standing research behind it and Artificial Intelligence (AI) has started to demonstrate how one can overcome various remaining problems, even those that are directly observation or experiment dependent. It sounds unusual and requires significant effort to understand how and why these procedures are working. 

Recent attempts to use such tools in physics (and not only in physics) have proven to be very promising, what hints towards the fact that very interesting results may be obtained when both fields are merged. Definitely, there are other interesting developments that could be here discussed, too. However, we will skip the discussion of developments in other fields since this is out of  the scope of the present work. \footnote{Given its aim, we omit here citing important papers devoted to many other sort of questions. Instead we address readers to the ArXiv and other web sources for comprehensive discussions on ML and AI in those other contexts.}

Our paper is aimed at the study of dark energy and Cosmology at large, and our goal is to produce a specific dark energy model, in a model-independent way, by using the advantages of a specific Machine Learning approach \cite{Bamba:2012cp} - \cite{INEnd} (see references therein for additional discussiosn on different developments concerning dark energy models and related problems). In particular, we will study a quintessence dark energy Universe when a Gaussian Process (GP) is involved (see for instance \cite{QDE} and \cite{QDE_1} covering some discussion about quintessence dark energy models). GPs provide interesting departures from standard reasoning in various fields. Their recent applications to Cosmology showed very interesting departures not reported previously \cite{GP_0}-\cite{GP_5} (and references therein for other GP applications in Cosmology). The reconstruction of $f(T)$ gravity from the expansion rate data, allowing to obtain very tight constraints on the model parameters of some popular $f(T)$ models is among them \cite{GP_0}. Moreover, a recent paper by two of the authors has shown how GPs can be used to tackle the Swampland Criteria for a  dark energy dominated Universe in a model-independent way \cite{GP_1}. In particular, it has been demonstrated there that the expansion rate data can be used, instead of assuming a specific form for the potential describing the quintessence dark energy, to tackle the Swampland Criteria. In other words, the whole analysis is based on the expansion rate data allowing to explore the features, which in some sense could be biased if a specific dark energy model were to be used. In this way, a hint indicating that the Swampland Criteria in its recent form is not suitable for a dark energy dominated Universe has been found. Among other interesting results, it was found that an effective theory being in the Swampland could (or not) end up  there. Moreover, starting out of the Swampland it is possible to end up either inside or outside of it (more details can be found in \cite{GP_1}). Having such interesting results in our hands, probably it would be possible in the near future to have unsuspected departures from the standard reasoning about the effective field theories (EFT) which is a promising task to be tackled down yet.  The fact that ML is designed to find the questions from the answers allows to hope that, in the near future,  interesting developments in this direction may arise. We would like to mention that there is another interesting approach, known as Bayesian Machine Learning, which we hope can eventually be very efficient in overcoming such limitations \cite{H0start_1, H0start_7, H0start_8} (see there how it can be used to tackle the $H_{0}$ tension problem). 

Now, let us come back to Ref. \cite{GP_0} to summarize what was done in that work. In the analysis, differently as in other works, no specific form for the potential describing the quintessence dark energy has been used. However, having a closer look at it, it is easy to see that the reconstruction of  the quintessence dark energy potential itself is possible too, in a model independent way, allowing also to obtain the constraints on the existing models. Moreover, it can be used to craft new models and gain some hints how, for instance, the $H_{0}$ tension problem can be solved in a quintessence dark-energy dominated Universe. In this way, we can indicate that the results of that paper  provide a unique possibility to treat quintessence dark energy models, that was not done in previous papers on the topic. Motivated by this possibility, given by the use of GPs,  the potential was reconstructed, in a model independent way  from available expansion rate data and, in addition, a new viable quintessence dark energy model was obtained. In particular, based on the mean of the obtained reconstructions, the potential $V(\phi) \sim \phi^{\lambda} \left [ 1 - sin^{n} (\beta \phi) \right ]$ has been proposed as a new form of quintessence dark energy---up to our best knowledge this potential has not been discussed anywhere previously. Other models were also considered, as $V(\phi) \sim \phi^{\lambda}$,  $V(\phi) \sim \phi^{\lambda} \left [ 1 - cos(\beta \phi^{n} ) \right ]$ and $V(\phi) \sim e^{-\lambda \phi}$ and values of the models parameters were estimated, indicating when they could be 1) viable and interesting for Cosmology, and 2) used to solve the $H_{0}$ tension problem. 

To end this section, we would like to mention also that, in our analysis here, we will use two kernels and consider three different cases for the value of the parameter $H_{0}$. The details will be discussed below and the whereabouts behind the method allowing to understand the results obtained will be presented in Sect. \ref{sec:BIBBL}. In this way, hints are given about the forms and constraints on the quintessence dark energy models that could be very useful in understanding how the $H_{0}$ tension problem could be alleviated. Moreover, up to our best knowledge, this is  the first  truly model independent reconstruction of the quintessence dark energy potential ever obtained. We do hope that these new results combined with the results discussed in Ref. \cite{GP_0} will be interesting for the community and lead to new developments in  future studies of quintessence dark energy models. There is also another interesting result which will be touched upon in the conclusion section.

This paper is organized as follows. The description of the GP is discussed in Sect.\ref{sec:BIBBL}. In the same section, we present the details allowing to reconstruct the potential. The main results are discussed in  Sect.\ref{sec:BEM}, which is followed by an analysis of their implications. Final conclusions of the analysis are given in Sect.\ref{sec:conc}.

\section{The method and the model}\label{sec:BIBBL}

The goal of this work is to provide a model-independent reconstruction of the quintessence dark energy potential by using a GP. We will present some details of how this can be achieved and on what  the involved assumptions are. We shall start from the background dynamics demonstrating what are the steps to follow to make the GP work, while some discussion on the GP itself will be presented at the end of this section. 

We consider General Relativitiy (GR) with the standard matter field in the presence of a quintessence field $\phi$, given by the following action ($8\pi G = c = 1$)
\begin{equation}
S = \int d^{4} x \sqrt{-g} \left( \frac{1}{2} R - \frac{1}{2} \partial_{\mu} \phi \partial^{\mu} \phi -V(\phi) \right ) + S_{m},
\end{equation}
where $\phi$ is the field, $V(\phi)$ the field's potential, $S_{m}$ corresponds to standard matter, while $R$ is the Ricci scalar. Moreover, it is well known that, when we consider a FRWL universe with 
\begin{equation}
ds^{2} = -dt^{2} + a(t)^{2} \sum_{i =1}^{3} (dx^{i})^{2},
\end{equation}
the dynamics of the scalar field's dark energy and dark matter are described by the equations 
\begin{equation}\label{eq:drhoPi}
\dot{\rho}_{\phi} + 3 H (\rho_{\phi} + P_{\phi}) = 0,
\end{equation}
\begin{equation}\label{eq:drhoDm}
\dot{\rho}_{dm} + 3 H \rho_{dm} = 0,
\end{equation}
with 
\begin{equation}\label{eq:F1}
H^{2} = \frac{1}{3} ( \rho_{\phi} +  \rho_{dm} ).
\end{equation}
In other words, Eqs. (\ref{eq:drhoPi}), (\ref{eq:drhoDm}) and (\ref{eq:F1})  describe the background dynamics. Furthermore, it is well known that $\rho_{\phi}$, $\rho_{dm}$ and $P = P_{\phi}$ are related to each other through the equation
\begin{equation}\label{eq:F2}
\dot{H} + H = -\frac{1}{6} (\rho_{\phi} + \rho_{dm} + 3 P_{\phi}).
\end{equation}
 On the other hand, assuming that the scalar field is spatially homogeneous, for its energy density and pressure we have
\begin{equation}\label{eq:rhoPhi}
\rho_{\phi} = \frac{1}{2} \dot{\phi} + V(\phi),
\end{equation}
and
\begin{equation}\label{eq:pPhi}
P_{\phi} = \frac{1}{2} \dot{\phi} - V(\phi),
\end{equation}
where the dot means derivative w.r.t the cosmic time, while $V(\phi)$ is the potential of the scalar field. In all equations  above $H = \dot{a}/a$ is the Hubble parameter. This is  well known, and also the important fact that the analysis of the background dynamics requires assuming the form of the potential $V(\phi)$; various forms for it  have been considered in the literature. 

Anyway, after some simple algebra one can see that, starting from Eqs. (\ref{eq:rhoPhi}) and(\ref{eq:pPhi}), it turns out that
\begin{equation}\label{eq:dphi2}
\dot{\phi}^{2} = \rho_{\phi} + P_{\phi},
\end{equation} 
while
\begin{equation}\label{eq:Vphi}
V(\phi) = \frac{\rho_{\phi} - P_{\phi}}{2}.
\end{equation}    
On the other hand, it is easy to see that from Eq. (\ref{eq:drhoDm}) we have $\rho_{dm} = 3 H_{0}^{2} \Omega_{0} (1+z)^{3}$, while from Eq. (\ref{eq:F1}) we can determine the energy density of the scalar field
\begin{equation}\label{eq:rhoH}
\rho_{\phi} = 3 H^{2} - 3 H_{0}^{2} \Omega_{0} (1+z)^{3},
\end{equation}
where $H_{0}$ is the value of the Hubble parameter at $z=0$ (e.g., at present; $z$ is the redshift). Now, we can use Eq. (\ref{eq:F2}) and, after some algebra,  
\begin{equation}\label{eq:PH}
P_{\phi} = 2(1+z)H H^{\prime} - 3 H^{2},
\end{equation}
where the prime denotes  derivative w.r.t. the redshift. Coming back to Eqs. (\ref{eq:dphi2}) and (\ref{eq:Vphi}), we see that Eqs. (\ref{eq:rhoH}) and (\ref{eq:PH}) allow to write down the form of the scalar field potential in terms of $H$ and $H^{\prime}$, as follows
\begin{equation}\label{eq:Vz}
V(z) = 3 H^{2} - H^{\prime} H (z+1)  - \frac{3}{2} H_{0}^{2} \Omega_{0} (z+1)^{3}.
\end{equation}
Moreover, it is possible to see that, for the field itself, we have
\begin{equation}\label{eq:phiz}
\left (\phi(z)^{\prime} \right) ^{2}  = \frac{2 H^{\prime} H - 3 H_{0}^2 \Omega_{0} (z+1)^2}{(1 + z)H^2 },
\end{equation}
what allows to perform an end-to-end reconstruction of $V(\phi)$, provided $H(z)$ and $H^{\prime}(z)$ are known. It should be noted that in Ref. \cite{GP_0} several other further steps have been taken too, in order to study the Swampland criteria; however, the discussion carried out here and the ensuing results, had never been discussed before.  

Now, it is time to make more transparent to the reader how can one obtain a model independent reconstruction of $V(\phi)$, from Eqs. (\ref{eq:Vz}) and (\ref{eq:phiz}). It is easy to see, to start, that Eqs. (\ref{eq:Vz}) and (\ref{eq:phiz}) allow to do this, if model independent reconstructions of $H(z)$ and $H^{\prime}(z)$ are provided. Namely, following Ref. \cite{GP_0}, we choose the GP to reconstruct $H$ and $H^{\prime}(z)$ from available expansion rate data (see Table \ref{tab:Table0}). Therefore the rest of this section is devoted to the presentation of some crucial aspects of GPs. To start, we recall that GPs are Bayesian state-of-the-art tools and that the key ingredient is the covariance function. In a nutshell, it is assumed that a GP prior governs the set of possible latent functions, and the likelihood of the latent function and observations shape this prior to produce posterior probabilistic estimates. The advantage of a GP is providing a full conditional statistical description of the predictions used to establish confidence intervals and to set hyper-parameters. For a given set of observations, it can infer the relation between independent and dependent variables. Moreover, GPs should be understood as distributions over functions, characterized by a mean function and a covariance matrix. Unfortunately, one disadvantage of the method is that the choice of the kernel is not a fixed process. Only well-designed data and the type of task to be tackled can indicate which kernel works better. A number of possible choices for the covariance function exist ---as squared exponential, polynomial, spline, etc., to mention a few. In other words it is always highly recommended to consider several kernel and compare the  results obtained, in order to be sure that the reconstruction has not been got by chance. This is very important and not treating this aspect very seriously can lead to misleading results, with bad consequences. In Cosmology we deal with relatively small datasets, therefore it is always possible to follow the reconstruction process allowing to significantly reduce the kernel numbers to be considered. This is one of the reasons that in Cosmology we usually meet studies involving only two or three kernels. In particular, in Cosmology one of the most actively used kernels is the squared exponential function
\begin{equation}\label{eq:kernel1}
k(x,x^{\prime}) = \sigma^{2}_{f}\exp\left(-\frac{(x-x^{\prime})^{2}}{2l^{2}} \right),
\end{equation}
where $\sigma_{f}$ and $l$ are parameters known as hyperparameters.  The $l$ parameter represents the correlation length along which the successive $f(x)$ values are correlated, while to control the variation in $f(x)$ relative to the mean of the process we need the $\sigma_{f}$ parameter. On the other hand, the squared exponential function is infinitely differentiable, which is a useful property in case of constructing higher-order derivatives. Recently, other kernels including  the so-called Matern ($\nu = 9/2$) covariance function
$$k_{M}(x,x^{\prime}) = \sigma^{2}_{f} \exp \left(-\frac{3|x-x^{\prime}|}{l} \right) $$
\begin{equation}\label{eq:kernel2}
\times \left[ 1+ \frac{3 |x-x^{\prime}|}{l} + \frac{27(x-x^{\prime})}{7l^{2}} + \frac{18|x-x^{\prime}|^{3}}{7l^{3}} + \frac{27 (x-x^{\prime})^{4}}{35 l^{4}}\right],
\end{equation}
 have been used for different purposes, too. Following this benchmark, we have also considered the squared exponential, Eq. (\ref{eq:kernel1}), and the Matern ($\nu = 9/2$), Eq. (\ref{eq:kernel2}), which allow eventually to understand 1) how they can affect the reconstruction of $V(\phi)$, and 2) how it can affect the constraints on $V(\phi)$ potentials existing in the literature. It should be mentioned that we have considered a god number of particular cases, but the reconstructions here presented are well enough to revisit all existing models. We will come back to this  in the next section, when we discuss the results obtained. 

Now, having closed the question of the kernel functions, let us discuss further about 1) the data and 2) the tools we use. In particular, the data used is the expansion rate values presented in Table \ref{tab:Table0}, consisting of  $30$-point samples of $H(z)$ deduced from the differential age method in addition to $10$-point samples obtained from the radial BAO method. In total we use $40$ data-points covering the $z\in [0, 2.4]$ redshift range. One interesting aspect of our analysis concerning the value of $H_{0}$ at $z=0$ to be mentioned here is about the adopted strategy. To wit, during the reconstruction, we consider three different cases: 1) $H_{0}$ is  estimated from the expansion rate data during the reconstruction of $H(z)$ and $H^{\prime}(z)$ , 2) $H_{0}$ is taken  from the Planck mission and the forms of $H(z)$ and $H^{\prime}(z)$ are reconstructed, and finally 3) $H_{0}$ is the one from the Hubble mission and then the forms of $H(z)$ and $H^{\prime}(z)$ are reconstructed. The reason for this, as it can be realized, is to see whether or not it is possible to find ways to solve or at least alleviate the $H_{0}$ tension problem. Ideally, using a model-independent reconstruction of the potential can indicate what are the shapes of the same allowing to solve the $H_{0}$ problem, if we follow the  strategy above. To this, we will come again in the next section. 

To end this section we need to mention that we use the publicly available package GaPP (Gaussian Processes in Python) developed by Seikel et al \cite{Seikel}. It is a very easy one to use and a very friendly package allowing to choose different covariance functions (new ones can be added easily, too). Moreover, the squared exponential function, Eq. (\ref{eq:kernel1}), is used in the code as a default option, while the Matern covariance function given by Eq. (\ref{eq:kernel2}), is already implemented. On the other hand, the code is very useful to combine different observational datasets, provided the proper relation between them is known. The package has been  often used, and more details about it, including a detailed description of the GP can be found in the references of our paper. In the next section we will describe our results thoroughly, which together with the discussion in the above section, will surely allow the readers to understand how the scheme of the reconstruction of the potential of the quintessence dark energy can be extended and applied on the other dark energy models, as phantom or tachyonic models.

\begin{table}[t]
  \centering
    \begin{tabular}{ |  l   l   l  |  l   l  l  | p{2cm} |}
    \hline
$z$ & $H(z)$ & $\sigma_{H}$ & $z$ & $H(z)$ & $\sigma_{H}$ \\
      \hline
$0.070$ & $69$ & $19.6$ & $0.4783$ & $80.9$ & $9$ \\
         
$0.090$ & $69$ & $12$ & $0.480$ & $97$ & $62$ \\
    
$0.120$ & $68.6$ & $26.2$ &  $0.593$ & $104$ & $13$  \\
 
$0.170$ & $83$ & $8$ & $0.680$ & $92$ & $8$  \\
      
$0.179$ & $75$ & $4$ &  $0.781$ & $105$ & $12$ \\
       
$0.199$ & $75$ & $5$ &  $0.875$ & $125$ & $17$ \\
     
$0.200$ & $72.9$ & $29.6$ &  $0.880$ & $90$ & $40$ \\
     
$0.270$ & $77$ & $14$ &  $0.900$ & $117$ & $23$ \\
       
$0.280$ & $88.8$ & $36.6$ &  $1.037$ & $154$ & $20$ \\
      
$0.352$ & $83$ & $14$ & $1.300$ & $168$ & $17$ \\
       
$0.3802$ & $83$ & $13.5$ &  $1.363$ & $160$ & $33.6$ \\
      
$0.400$ & $95$ & $17$ & $1.4307$ & $177$ & $18$ \\

$0.4004$ & $77$ & $10.2$ & $1.530$ & $140$ & $14$ \\
     
$0.4247$ & $87.1$ & $11.1$ & $1.750$ & $202$ & $40$ \\
     
$0.44497$ & $92.8$ & $12.9$ & $1.965$ & $186.5$ & $50.4$ \\

$$ & $$ & $$ & $$ & $$ & $$\\ 

$0.24$ & $79.69$ & $2.65$ & $0.60$ & $87.9$ & $6.1$ \\
$0.35$ & $84.4$ & $7$ &  $0.73$ & $97.3$ & $7.0$ \\
$0.43$ & $86.45$ & $3.68$ &  $2.30$ & $224$ & $8$ \\
$0.44$ & $82.6$ & $7.8$ &  $2.34$ & $222$ & $7$ \\
$0.57$ & $92.4$ & $4.5$ &  $2.36$ & $226$ & $8$ \\ 
          \hline
    \end{tabular}
    \vspace{5mm}
\caption{$H(z)$ and its uncertainty $\sigma_{H}$  in  units of km s$^{-1}$ Mpc$^{-1}$. The upper panel consists of thirty samples deduced from the differential age method. The lower panel corresponds to ten samples obtained from the radial BAO method. The table is according to \cite{GP_0} (see also references therein for details).}
  \label{tab:Table0}
\end{table}

\section{Results and discussion}\label{sec:BEM}

\begin{figure}[t!]
 \begin{center}$
 \begin{array}{cccc}
\includegraphics[width=80 mm]{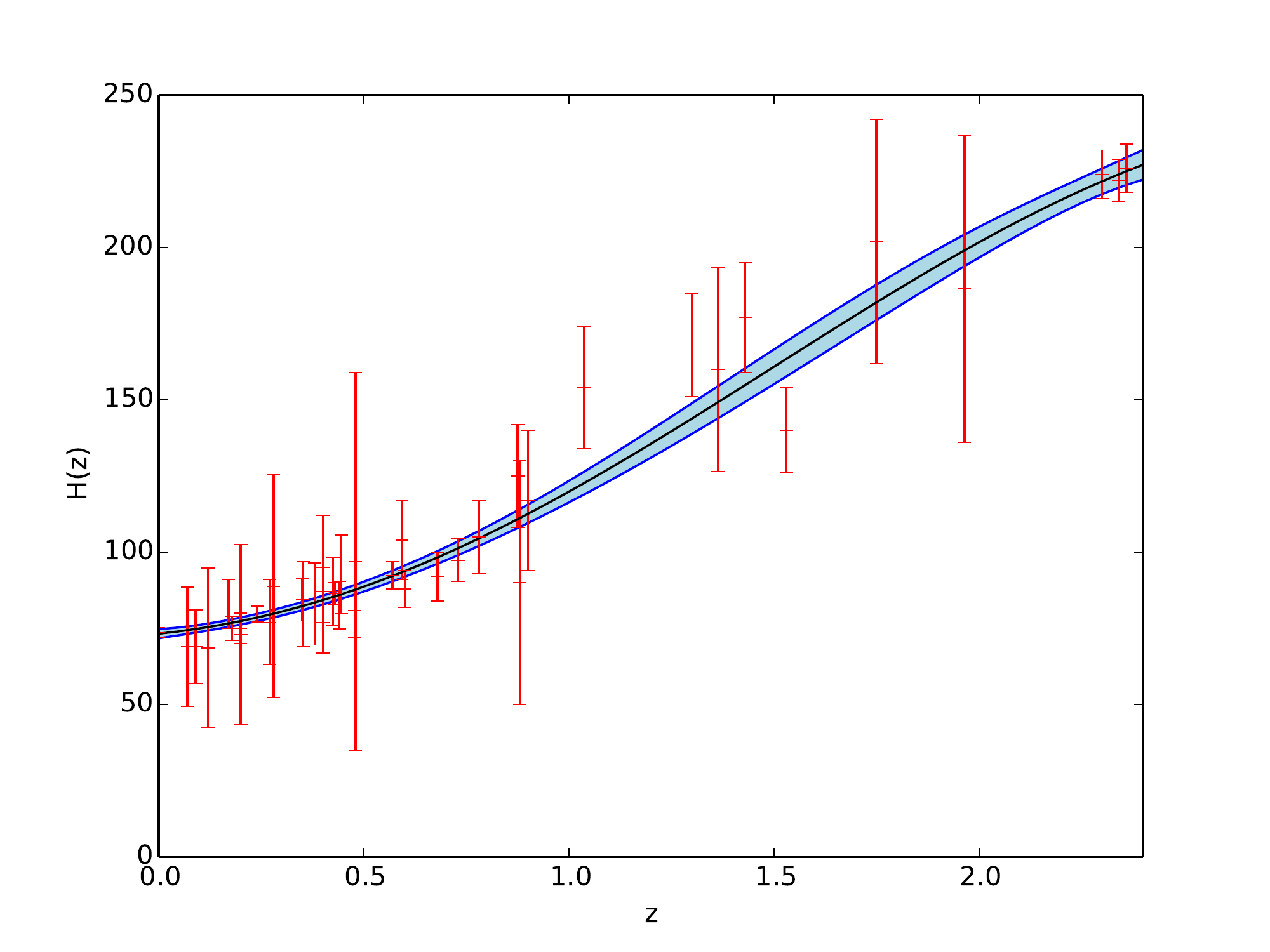}&&
\includegraphics[width=80 mm]{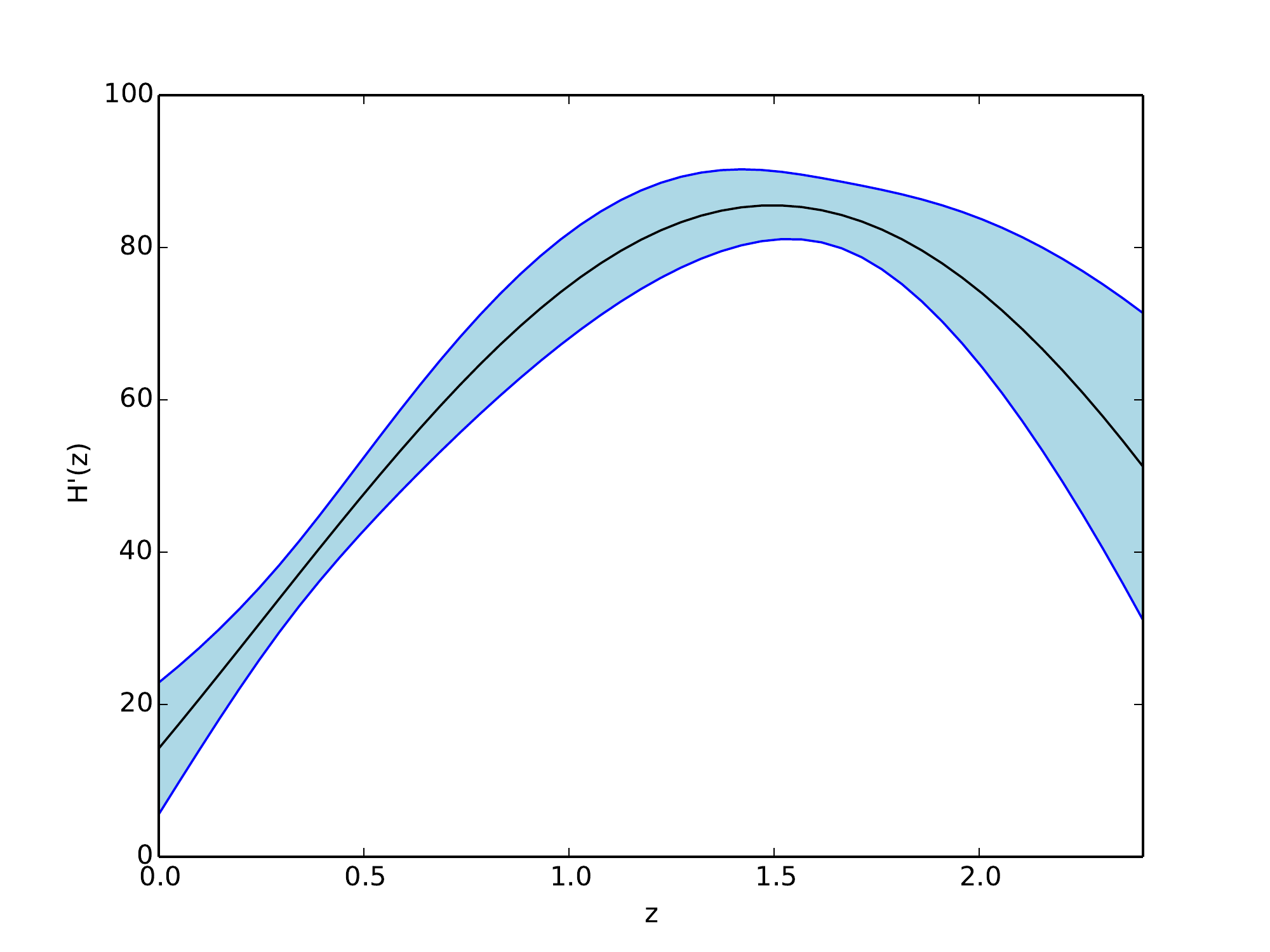}\\
 \end{array}$
 \end{center}
\caption{GP reconstruction of $H(z)$ and $H^{\prime}(z)$ for the 40-point sample deduced from the differential age method, with the additional 10-point sample obtained from the radial BAO method, when $H_{0} = 73.52 \pm 1.62$ km s$^{-1}$ Mpc$^{-1}$ reported by the Hubble mission. The $^{\prime}$ means derivative with respect to the redshift $z$.}
 \label{fig:Fig1}
\end{figure}

In this section we present and discuss our results. They can be split into three different cases, corresponding to the reconstruction when: 1) $H_{0}$ is  estimated from a GP, 2) $H_{0}$ is fixed to the value estimated from the Planck mission results and the reconstruction of $H(z)$ and $H^{\prime}(z)$ is performed, and 3) $H_{0}$ is fixed to the value estimated from the Hubble mission and then the reconstruction of $H(z)$ and $H^{\prime}(z)$ is performed. We have already mentioned that, in this way, we can get a hint on when the $H_{0}$ tension problem could be solved and on what are the constraints on some explicitly given model parameters in order to achieve the solution. Moreover, we  should mention again that, in our analysis, we use two kernel functions, given namely by Eqs. (\ref{eq:kernel1}) and (\ref{eq:kernel2}). We are interested in a model-independent reconstruction of the quintessence dark energy potential and we use the expansion rate data and the GP to reconstruct $H(z)$ and $H^{\prime}(z)$  in Eqs. (\ref{eq:Vz}) and (\ref{eq:phiz}). In Fig.(\ref{fig:Fig1}) a reconstruction of the functions $H(z)$ and $H^{\prime}(z)$ for the squared exponential function, Eq. (\ref{eq:kernel1}), has been used in a GP, assuming that $H_{0} = 73.52 \pm 1.62$ km s$^{-1}$ Mpc$^{-1}$. The functions $H(z)$ and $H^{\prime}(z)$ corresponding to other cases can be reconstructed in a similar way. A crucial point, not discussed in the previous section, is how to deal with Eq. (\ref{eq:phiz}) since eventually we will reconstruct $V(\phi)$. This problem is the simplest one which can be tackled if we take into account that 
\begin{equation}\label{eq:deltaphi}
\phi(z_{i})^{\prime} \approx \frac{ \phi(z_{i} + \Delta z)  - \phi(z_{i}) }{\Delta z},
\end{equation}  
where $\Delta z = z_{i+1} - z_{i}$ with $z_{i}$ correspond to the redshifts where $H(z)$ and $H^{\prime}(z)$ have been reconstructed. It is clear, that Eqs. (\ref{eq:Vz}),  (\ref{eq:phiz}) and (\ref{eq:deltaphi}) allow to perform the model independent reconstruction of the potential $V(\phi)$  describing  quintessence dark energy in our Universe. For the sake of convenience, we divide our results in three subsections. 

\subsection{$V(\phi)$ reconstruction when $H_{0}$ is not fixed}

The first case corresponds to the reconstruction when  $H_{0}$ is not fixed. In this case, using GP a and the expansion rate data, as in Table \ref{tab:Table0}, we first estimate $H_{0}$ during the reconstruction process, which is found to be $H_{0} = 71.286 \pm 3.743$ km s$^{-1}$ Mpc$^{-1}$ when the kernel is given by Eq. (\ref{eq:kernel1}), while, when the kernel is given by Eq. (\ref{eq:kernel2}), we found that $H_{0} = 71.196 \pm 3.867$  km s$^{-1}$ Mpc$^{-1}$. Then using the reconstructed $H(z)$ and $H^{\prime}(z)$ in Eqs. (\ref{eq:Vz}) and (\ref{eq:phiz}) (combined with Eq. (\ref{eq:deltaphi})) we are able to finish the model independent recunstruction of $V(\phi)$. Omitting other non-relevant technical details, we refer the reader to Fig. (\ref{fig:Fig2}), which depicts the model-independent reconstructed forms of the potential $V(z)$ and field $\phi(z)$. The reader may have already noted that the estimated errors for $H_{0}$  are significantly larger than those from the Planck data and those from the Hubble mission. It should be mentioned that, as a consequences of these upper and lower bounds on $V(\phi)$, this case will significantly differ from the other two cases. In general, this can have a strong impact on the model parameter constraints and affect the viable model selection. Obviously, in general this can affect the early time behavior of a given quintessence dark energy model.

We now elaborate on our reconstruction results. At first glance, the reconstruction has been successful. However, in order to understand and validate the corresponding results, we need to have a look at another physical quantity describing our model and that has been reconstructed, too. Since it could be, as  is demonstrated in many studies, that the model under consideration, due to the quality of the observational data, is bound to be rejected or it is valid up to some redshift value, only. Therefore, it is important to understand up to what extent we can believe in the model validity in our case. In general, non-validation of the reconstruction results can lead to wrong interpretations and cause misunderstanding of the underlying physics. In our case, a non-proper treatment of the model validity issue can lead to a number of different problems, including the validation of a wrong form of the potential $V(\phi)$. The physical quantity we choose for this purposes is  $\Omega_{de} = \rho_{\phi}/3H^{2}$, which at $z=0$ has actually been used to estimate $\phi(z=0)$, too. 

The results of the reconstruction of the $\Omega_{\phi}$ can be found in Fig. (\ref{fig:FigA_2}) for both kernels given by Eqs. (\ref{eq:kernel1}) and (\ref{eq:kernel2}), respectively. Indeed, we see that the reconstruction was successful up to a certain redshift, indicating that for higher values the model should be rejected, since the lower $2\sigma$ bound of $\Omega_{de}$ is negative. It is important to mention that the GP give the statistical explanation of the results, and considering only the mean to decide whether or not something is working is not a correct procedure. In order to have a proper understanding, we need to consider the whole picture, which in our case indicates that the reconstruction of $V(z)$ and $\phi(z)$ is acceptable up to some redshift. Having this in mind, we continued the study and, using the means of the reconstructed $V(z)$ and $\phi(z)$, we have directly reconstructed the mean of $V(\phi)$. 

\begin{figure}[h!]
 \begin{center}$
 \begin{array}{cccc}
\includegraphics[width=85 mm]{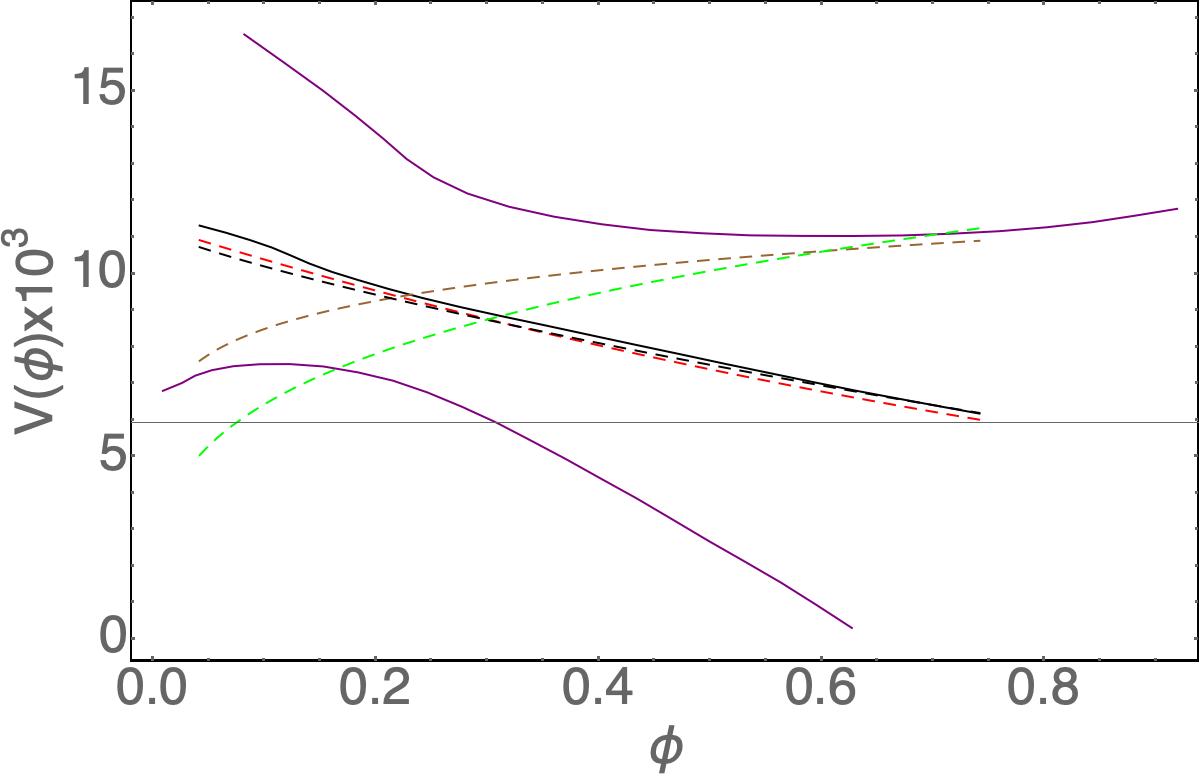}&&
\includegraphics[width=85 mm]{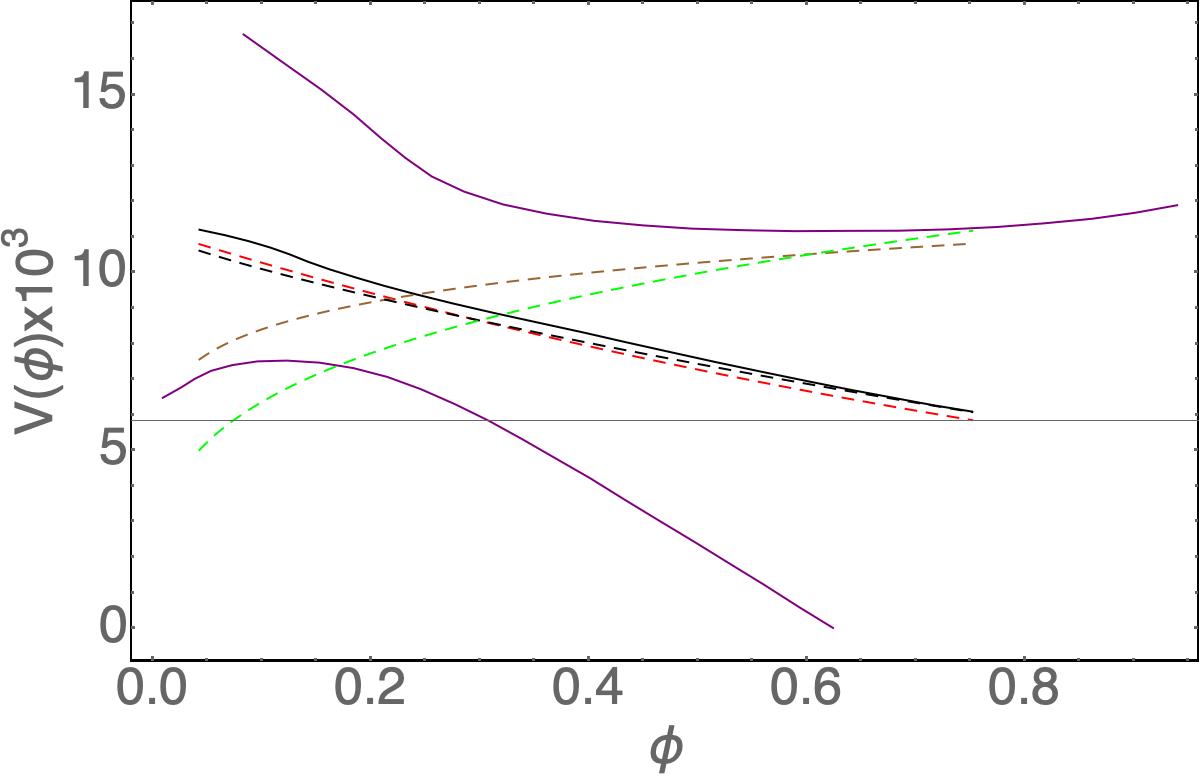}\\
 \end{array}$
 \end{center}
\caption{Reconstructed $V(\phi)$ for the case when $H_{0} = 71.286 \pm 3.743$ km s$^{-1}$ Mpc$^{-1}$ has been estimated by a GP using the expansion rate data presented in Table \ref{tab:Table0}. The black curve corresponds to the mean of the reconstructed $V(\phi)$ model obtained from the reconstructed means of the functions $V(z)$ and $\phi(z)$. The lower and upper $2\sigma$ error bounds of both reconstructed functions $V(z)$ and $\phi(z)$ have been used to determine possible maximum error bounds for $V(\phi)$ (purple curves). The dashed red curve represents the quintessence dark energy model with potential $V(\phi) \sim e^{-\lambda \phi}$. The dashed brown, dashed black and dashed green curves correspond to the quintessence dark energy model with $V(\phi) \sim \phi^{\lambda}$, $V(\phi) \sim \phi^{\lambda} \left [ 1 - sin^{n} (\beta \phi) \right ] $ and $V(\phi) \sim \phi^{\lambda} \left [ 1 - cos(\beta \phi^{n} ) \right ]$, respectively. The left hand side plot is the result of the reconstruction when the kernel is given by Eq. (\ref{eq:kernel1}), while the right hand side one stands for the case with kernel given by Eq. (\ref{eq:kernel2}).}
 \label{fig:Fig2_1}
\end{figure}

On the other hand, using the lower and upper $2\sigma$ error bounds of both the $V(z)$ and $\phi(z)$ reconstructed functions, we have determined possible maximum errors for $V(\phi)$ allowing to complete our task, which was to obtain a model independent reconstruction of $V(\phi)$ describing  quintessence dark energy as the driving force of our expanding Universe. The result can be found in Fig. (\ref{fig:Fig2_1}), where the left hand side plot represents the reconstruction result when the kernel is given by Eq. (\ref{eq:kernel1}), while the right hand side one stands for the case with kernel given by Eq. (\ref{eq:kernel2}), respectively. 

Fig. (\ref{fig:Fig2_1}) shows how different models can be compared and constrained using the reconstruction. In particular, from our results here, namely the lhs plot of Fig.(\ref{fig:Fig2_1}), when the kernel is given by Eq.(\ref{eq:kernel1}), we clearly see why the quintessence dark energy model with $V(\phi) \sim e^{-\lambda \phi}$(red dashed curve, for instance, with  $\lambda = 0.854$) is among the most successful ones, and our analysis explains why this particular model has captured such a lot of attention in the literature. On the other hand, we also see that the quintessence dark energy model with $V(\phi) \sim \phi^{\lambda} \left [ 1 - cos(\beta \phi^{n} ) \right ]$ (dashed green curve with $\beta = 1.65$, $n=0.05$ and $\lambda = 0.2$, for instance) will not work very well and there is a hint that it should be rejected. Additionally, according to the existing expansion rate data, the model with potential $V(\phi) \sim \phi^{\lambda}$  (dashed brown curve with $\lambda = 0.125$) should be kept, and it will work better than the model with $V(\phi) \sim \phi^{\lambda} \left [ 1 - cos(\beta \phi^{n} ) \right ]$. This is inferred from the dashed brown curves on both plots of Fig.(\ref{fig:Fig2_1}). 

It is made clear from the provided discussion that any given model can be analysed, and that proper constraints on the parameters can be found. Up to this moment we have discussed some of the already existing models and, there, our results coming from the model-independent analysis give just an enhanced understanding of why some work and why others should be rejected. Going one step further, it should be now possible to see why some dark energy models with specific potentials will not work for solving the $H_{0}$ tension problem. Moreover, in the next subsection we will discuss how drastically the constraints on the model parameters should be changed in order to be suitable to solve the $H_{0}$ tension problem. 

Before ending this one, and based on the reconstructed results, we suggest a new form for the potential to describe a viable quintessence dark energy model. For what we know, this specific model has not been considered before. The potential has the form
\begin{equation}
V(\phi) \sim \phi^{\lambda} \left [ 1 - sin^{n} (\beta \phi) \right ],
\end{equation}
where $\lambda$, $n$ and $\beta$ are free parameters  to be fitted. The dashed black curve of Fig.(\ref{fig:Fig2_1}) corresponds to this model; it is one-to-one in mimicking the reconstructed mean behavior with $\lambda = 0.001$, $\beta = 0.5$ and $n=0.5$, when the kernel is given by  Eq.(\ref{eq:kernel1}). 
To finish, we should  mention that the consideration of the kernel, Eq.(\ref{eq:kernel2}), will introduce changes only slightly affecting the above discussed numerical values of the parameters. However, the general picture and the conclusions drawn remain unchanged.

\subsection{$V(\phi)$ reconstruction when $H_{0} = 67.40 \pm 0.5$ km s$^{-1}$ Mpc$^{-1}$}\label{Planck}

In this subsection we discuss the case when the reconstruction of $H(z)$ and $H^{\prime}(z)$ has been performed when a specific value for $H_{0}$  has been fixed in advance. Differently to the previous case, now the reconstruction is based on $41$ data points. We will later see that this can affect our perception of the situation and can be moreover useful to understand why the constraints on quintessence dark energy models discussed in the recent literature may be so different, and what is the path to follow to solve the $H_{0}$ tension problem. To be more precise let us indicate that, in this case, instead of estimating the value of $H_{0}$, we use the $H_{0}$ reported by the Planck mission. Similar to the previous case, $\Omega_{de} = \rho_{\phi}/3H^{2}$ has been considered again allowing to determine the redshift range where the reconstruction is valid. In particular, we found that when the squared exponent kernel given by Eq.(\ref{eq:kernel1}) is  considered then the $z\in [0,2)$ redshift range provides physically acceptable reconstruction of the functions $V(z)$ and $\phi(z)$.  Moreover, when we consider the Matern ($\nu = 9/2$) kernel, Eq. (\ref{eq:kernel2}), then  $z\in [0,1.91)$ is the redshift range providing physically acceptable reconstructions of $V(z)$ and $\phi(z)$ (see Fig. (\ref{fig:FigA_3})). On the other hand, from the top panel of Fig. (\ref{fig:Fig3}) we realize the reconstructed forms of the potential $V(z)$ and the field $\phi(z)$, when the kernel is given by  Eq. (\ref{eq:kernel1}). Complementary, the reconstruction results when the kernel is given by  Eq. (\ref{eq:kernel2}) can be found on the bottom panel of  Fig. (\ref{fig:Fig3}).

\begin{figure}[h!]
 \begin{center}$
 \begin{array}{cccc}
\includegraphics[width=85 mm]{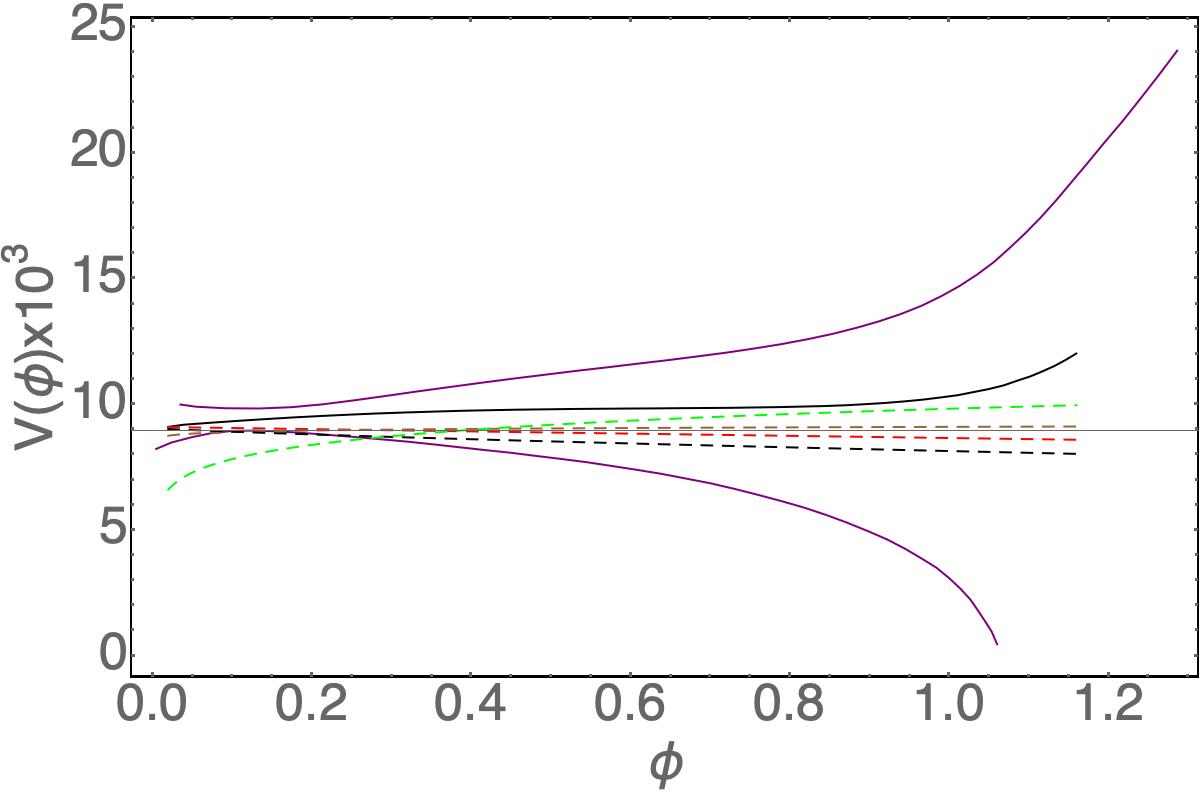}&&
\includegraphics[width=85 mm]{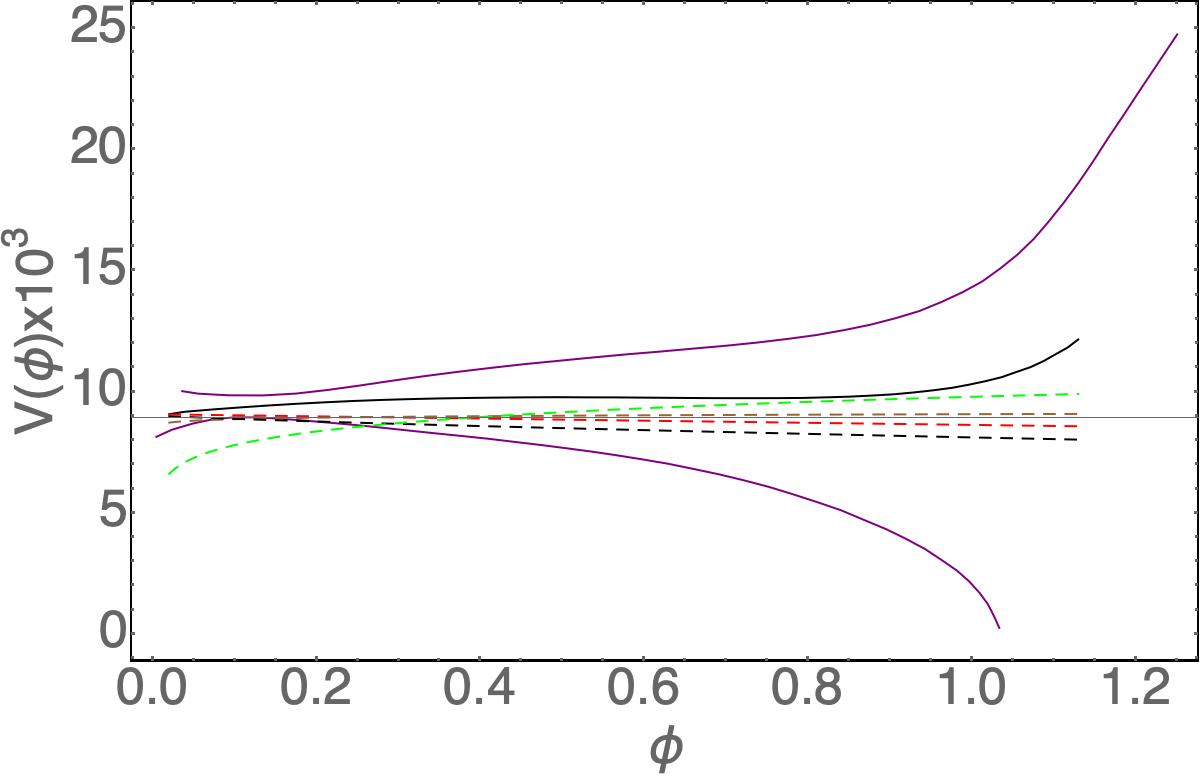}\\
 \end{array}$
 \end{center}
\caption{Reconstructed $V(\phi)$ for the expansion rate data  in Table\ref{tab:Table0} with $H_{0} = 67.40 \pm 0.5$ km s$^{-1}$ Mpc$^{-1}$, as reported by the Planck mission, which has been used for the reconstruction. The black curve represent the mean of the reconstructed $V(\phi)$ model obtained from the reconstructed means of $V(z)$ and $\phi(z)$. The lower and upper $2\sigma$ error bounds for both reconstructed functions, $V(z)$ and $\phi(z)$, have been used to determine possible maximum error bounds for $V(\phi)$ (purple curves). The dashed red curve corresponds to the quintessence dark energy model with potential $V(\phi) \sim e^{-\lambda \phi}$. Dashed brown, dashed black and dashed green curves represent the quintessence dark energy model with $V(\phi) \sim \phi^{\lambda}$, $V(\phi) \sim \phi^{\lambda} \left [ 1 - sin^{n} (\beta \phi) \right ] $ and $V(\phi) \sim \phi^{\lambda} \left [ 1 - cos(\beta \phi^{n} ) \right ]$, respectively. The left hand side plot depicts the reconstruction result when the kernel is given by Eq. (\ref{eq:kernel1}), while the right hand side one stands for the case with kernel given by Eq. (\ref{eq:kernel2}). }
 \label{fig:Fig3_1}
\end{figure}

To reconstruct the mean of $V(\phi)$ potential we have used the means of the reconstructed $V(z)$ and $\phi(z)$. Moreover, using the lower and upper $2\sigma$ error bounds of both reconstructed functions, $V(z)$ and $\phi(z)$,  we have determined possible maximal errors for $V(\phi)$ (black curve in Fig.(\ref{fig:Fig3_1})), what allowed us to complete the model independent reconstruction of the $V(\phi)$ describing the quintessence dark energy in our expanding Universe. The reconstruction results can be found in Fig.(\ref{fig:Fig3_1}), where the left hand side plot represents the reconstruction result when the kernel is given by Eq.(\ref{eq:kernel1}), while the right hand side one corresponds to the case with kernel given by Eq.(\ref{eq:kernel2}). 

Visual comparison of the results presented in Fig. (\ref{fig:Fig2_1}) and Fig. (\ref{fig:Fig3_1}) already points out huge differences. In particular, just  comparing the mean of the reconstruction with $V(\phi) \sim e^{-\lambda \phi}$, we conclude that the model with $\lambda = 0.05$ should be preferred for cosmological applications. On the other hand,  with $\lambda = 0.01$ the model with the potential $V(\phi) \sim \phi^{\lambda}$ may be highly recommended for doing cosmology. Moreover, we also conclude that, with $\beta = 0.05$, $n=0.75$ and $\lambda = 0.001$, the model with potential $V(\phi) \sim \phi^{\lambda} \left [ 1 - sin^{n} (\beta \phi) \right ]$  is favored for doing cosmology. Finally, the model with  $V(\phi) \sim \phi^{\lambda} \left [ 1 - cos(\beta \phi^{n} ) \right ]$  can be recommended too, if $\lambda = 0.02$, $\beta = 1.65$ and $n=0.05$. 

In all these  examples, the kernel was given by Eq. (\ref{eq:kernel1}). Our analysis using the kernel of Eq. (\ref{eq:kernel2}) shows that similar recommendations  can be done. However, we should note that the reconstruction indicates that here we will have tighter constraints on the model parameters than in the previous case. However, the most relevant aspect revealed from the reconstruction is that the early time behavior of the models can change significantly. This is a hint indicating that the $H_{0}$ tension is not just a result of playing with numbers. It is more profound than this, namely a problem related with the physics and corresponding considerations. This should be made more clear in the next subsection, where we will present the results corresponding to the reconstruction  based on the value of $H_{0}$ reported by the Hubble mission.

\subsection{$V(\phi)$ reconstruction when $H_{0} = 73.52 \pm 1.62$ km s$^{-1}$ Mpc$^{-1}$}

\begin{figure}[h!]
 \begin{center}$
 \begin{array}{cccc}
\includegraphics[width=85 mm]{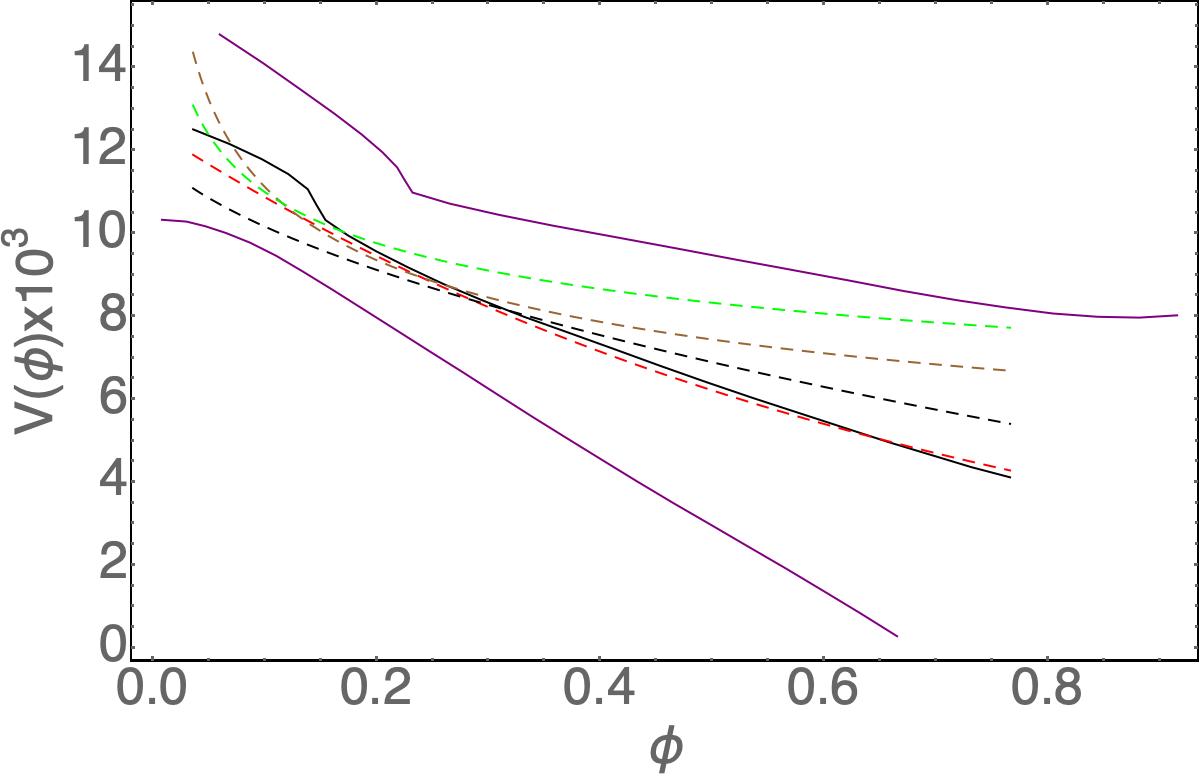}&&
\includegraphics[width=85 mm]{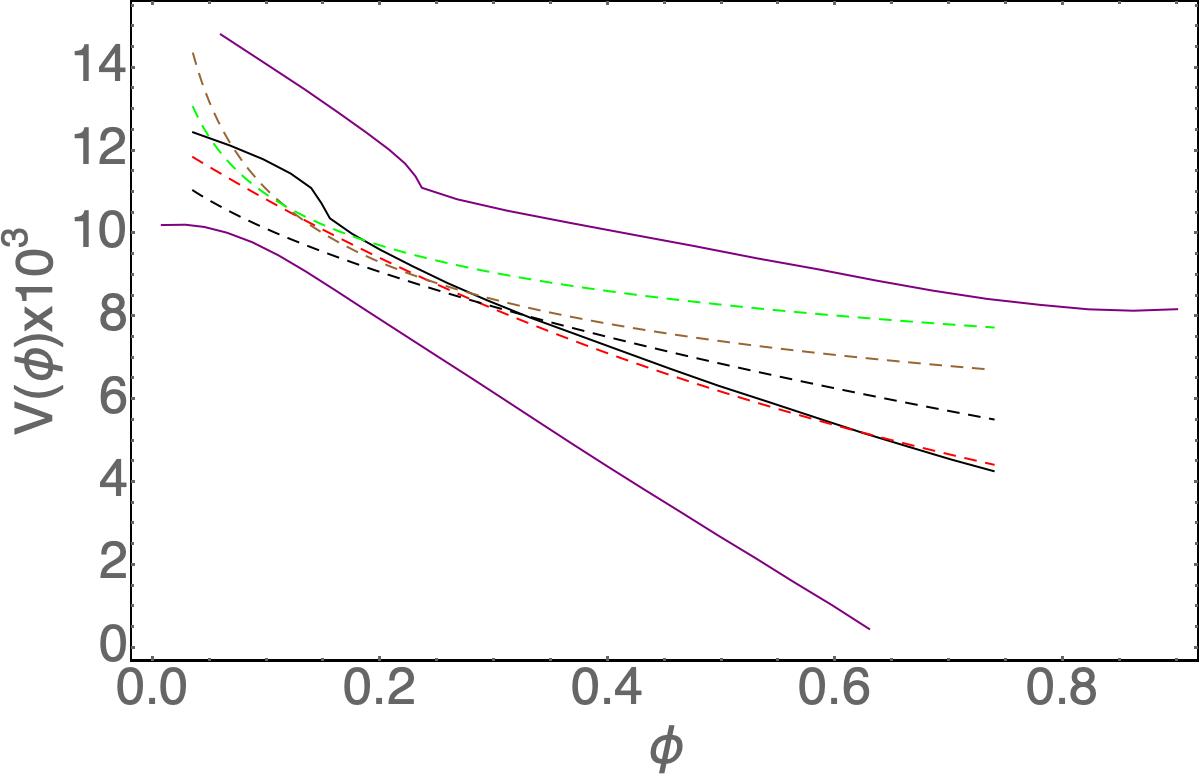}\\
 \end{array}$
 \end{center}
\caption{Reconstructed $V(\phi)$ for the  expansion rate data listed in Table\ref{tab:Table0} with $H_{0} = 73.52 \pm 1.62$ km s$^{-1}$ Mpc$^{-1}$, as reported by the Hubble mission, which have been used for the reconstruction. The black curve represents the mean of the reconstructed $V(\phi)$ model obtained from the reconstructed means of $V(z)$ and $\phi(z)$. The lower and upper $2\sigma$ error bounds of both reconstructed functions, $V(z)$ and $\phi(z)$,  have been used to determine possible maximum error bounds for $V(\phi)$(purple curves). The dashed red curve corresponds to the quintessence dark energy model with potential $V(\phi) \sim e^{-\lambda \phi}$.  Dashed brown, dashed black and dashed green curves correspond to the quintessence dark energy model with $V(\phi) \sim \phi^{\lambda}$, $V(\phi) \sim \phi^{\lambda} \left [ 1 - sin^{n} (\beta \phi) \right ] $ and $V(\phi) \sim \phi^{\lambda} \left [ 1 - cos(\beta \phi^{n} ) \right ]$, respectively. The left had side plot depicts the reconstruction result when the kernel is given by Eq. (\ref{eq:kernel1}), while the right hand side one stands for the case with the kernel of Eq. (\ref{eq:kernel2}).}
 \label{fig:Fig4_1}
\end{figure}

The last case to be discussed here is when  $H_{0}$ is  fixed to the value $H_{0} = 73.52 \pm 1.62$ km s$^{-1}$ Mpc$^{-1}$ reported by the Hubble mission. This means that, similarly to the second case, here the reconstruction of $H(z)$ and $H^{\prime}(z)$ will be also based on $41$ data points. The reconstructed $V(z)$ and $\phi(z)$ for $z\in[0,2.4]$ can be found in Fig. (\ref{fig:Fig4}). The  reconstruction there, has been obtained for the squared exponent kernel given by Eq.(\ref{eq:kernel1}). The final results of the reconstruction are depicted in Fig. (\ref{fig:Fig4_1}).

Our analysis based on these reconstructions shows when the resulting potentials can be recommended for doing Cosmology. In particular, we found that according to the mean of the $V(\phi)$ reconstruction, the model with $V(\phi) \sim e^{-\lambda \phi}$ and $\lambda = 1.4$ is expected to be useful for cosmological applications. On the other hand, the model with potential $V(\phi) \sim \phi^{\lambda}$  can be highly recommended, provided $\lambda = - 0.25$. Moreover, we found also that, with $\beta = 0.5$, $n=0.576$ and $\lambda = 0.005$, the model with potential $V(\phi) \sim \phi^{\lambda} \left [ 1 - sin^{n} (\beta \phi) \right ]$  is also favored for cosmological applications. Finally, the model with  $V(\phi) \sim \phi^{\lambda} \left [ 1 - cos(\beta \phi^{n} ) \right ]$ is also useful, provided $\lambda = -0.25$, $\beta = 1.75$ and $n=0.05$. In all these examples the kernel was given by Eq. (\ref{eq:kernel1}). An analysis using the kernel of Eq. (\ref{eq:kernel2}) shows that similar recommendations can be done, too. Moreover, the reconstruction results indicate that here the constraints on the model parameters will not be tighter than  in the case discussed in Sect. (\ref{Planck}). On the other hand, the early time behavior of the models could change significantly. The reconstruction of the $\Omega_{de} = \rho_{\phi}/3H^{2}$ has been considered again allowing to determine the redshift range where the reconstruction is valid (Fig. (\ref{fig:FigA_4})).

To end this subsection let us mention that the values of the  parameters for the models presented above give a hint on when the $H_{0}$ tension problem can be solved, in a quintessence dark energy dominated Universe, when one of the forms for the potential discussed above is used. Let us  mention again that reconstructions here obtained  are totally model independent and based on the expansion rate data, and that they can be used either to craft new models or either to constraint  already existing ones.

\section{Conclusions}\label{sec:conc}

In this paper, we have used GPs and available expansion rate data with the aim to reconstruct the functional form of the potential better representing quintessence dark energy. There are various and important open questions about dark energy physics and the challenge of answering them is usually undertaken using model-dependent approaches. The quintessence dark energy paradigm is among the most often discussed models. In there, the form of the potential field is chosen manually, in a sort of phenomenological approach, mainly aimed to reproduce the observational data. We notice that the same is correct for other dark energy models and that mainly phenomenology based motivations have been put forward to craft viable dark energy models. 

As an alternative to all these previously carried out studies, we here describe in detail the whereabouts of a model-independent reconstruction of the potential. Moreover, the results of the reconstruction  can be used to build new potentials, and  to constrain the free parameters. Starting from very basic assumptions about the background dynamics, we have demonstrated that the potential and the field itself can be expressed in terms of $H(z)$ and $H^{\prime}(z)$, which can be reconstructed in a model independent way from the expansion rate data using a GP. GPs are among a number of very useful  Machine Learning tools intensively used in very different areas, among them in Cosmology. 

The main issue with this approach is to specify the form of the kernel function that needs to be chosen in order to be able to complete the reconstruction. In this regard, the literature contains various interesting discussions, which can be summarized saying namely that,  for a given data and task, it is better just to use several kernels and compare the results. This is, in our opinion, an optimal solution that it can be time consuming; however, it is judicious to follow this approach and make sure that the hints and the results obtained  have a value and have not been got by  chance. 

In our work, the quintessence dark energy potential and the corresponding field  have been reconstructed for three different cases:  1) when $H_{0}$ has been estimated from the GP reconstruction of the functions  $H(z)$ and $H^{\prime}(z)$, based on an existing $40$-point expansion rate dataset; 2) when  $H_{0}$ is fixed to the value estimated from the Planck mission and then the reconstruction of $H(z)$ and $H^{\prime}(z)$ is performed; 3) when $H_{0}$ is fixed to the value estimated from the Hubble mission results and then the reconstruction of $H(z)$ and $H^{\prime}(z)$ is performed. In this way we get a hint on when the $H_{0}$ tension problem could be reasonably solved, for instance, when the model with $V(\phi) \sim e^{-\lambda \phi}$ is considered. We have studied other models, too, and also found a new potential, $V(\phi) \sim \phi^{\lambda} \left [ 1 - sin^{n} (\beta \phi) \right ] $, which can be recommended to be used in Cosmology. This is actually a genuinely new discovery made in this paper. An in depth study of this new potential has been left for a future publication.  

We need to mention that, in our analysis, we have used the two kernel functions given by Eqs. (\ref{eq:kernel1}) and (\ref{eq:kernel2}), and found slight changes (detailed above), as compared with the results discussed before, but those are not so important. However, we would like to discuss another result we have at this point. First, concerning the  possible constraints on the $\lambda$ parameter of the Swampland $V(\phi) \sim e^{-\lambda \phi}$ potential, which was discussed in Ref. \cite{Lavinia}. Without going deep into the discussion of Ref. \cite{Lavinia}, we learned that, with future surveys, we should expect fundamental observational limitations to lowering $\lambda$ to $\lambda < 0.1$ supporting the standard model. Now the question is: What we have learned with our method? In order to understand this, let us summarize what we  obtained, namely the preferred values for the parameters: 1) $\lambda = 0.854$ when $H_{0}$ is not fixed, 2) $\lambda = 0.01$ when $H_{0} = 67.40 \pm 0.5$ km s$^{-1}$ Mpc$^{-1}$, and finally 3) when $H_{0} = 73.52 \pm 1.62$ km s$^{-1}$ Mpc$^{-1}$ we got $\lambda = 1.4$. In all cases we have just used the reconstruction mean, and the last one means that, if we use lower error bounds of the reconstruction, we will reduce the estimated values too. 

Anyhow, the important message we wish to transmit to the reader is: 1) that the $40$ data point expansion rate data already contains the information that can come from future surveys; 2) the great importance of the tool we use to extract information from data. Moreover, our  constraints on the parameter $\lambda$  when $H_{0} = 67.40 \pm 0.5$ km s$^{-1}$ indicate that any other estimation closer to the estimation obtained here definitely supports the $\Lambda$CDM standard model. In all other cases we can claim that the $\Lambda$CDM theory might be challenged. This is an indication that the $H_{0}$ tension problem is not a game on statistics only. 

Of course, there are other various questions to be studied yet, which have been left for further consideration.  In particular, to continue using GPs and other Machine Learning algorithms involving other datasets for model independent reconstruction or pattern learning that can be used for similar estimations and reconstructions. More specifically, it would be interesting to see if the recently discovered constraints would be challenged in those cases, and what the consequences of this could be.

\section*{Acknowledgements}

This work has been partially supported by MICINN (Spain), project PID2019-104397GB-I00, of the Spanish State Research Agency program AEI/10.13039/501100011033, by the Catalan Government, AGAUR project 2017-SGR-247, and by the program Unidad de Excelencia María de Maeztu CEX2020-001058-M. This paper was supported by the Ministry of Education and Science  of the Republic of Kazakhstan, grant AP08052034.

\begin{thebibliography}{1}


\bibitem{H01}
N. Aghanim et al, Astron. Astrophys. 641 (2020) A6.  
 
 \bibitem{H02}
 A. G. Riess et al, Astrophys. J. 861 (2) (2018) 126. 
 
 \bibitem{H0Start}
G. S. Sharov, E. S. Sinyakov, arxiv:2002.03599.

\bibitem{H0start_1}
E. Elizalde et al, arXiv:2104.01077.

\bibitem{H0start_2}
M. Braglia et al, arxiv:2004.1116.

\bibitem{H0start_3}
W. L. Kimmy Wu et al, arxiv: 2004.10207.

\bibitem{H0start_4}
G. Alestas et al, arxiv:2004.08363.

\bibitem{H0start_5}
D. Wang, D. Mota, arxiv:2003.10095.

\bibitem{H0start_6}
J. Sakstein, M. Trodden, Phys.Rev.Lett. 124, 161301 (2020).

\bibitem{H0start_7}
E. Elizalde et al, arXiv:2006.12913.

\bibitem{H0start_8}
E. Elizalde et al, Phys.Rev.D 102, 123501 (2020).

\bibitem{H0start_9}
M. H. P. M. van Putten, arxiv:1707.02588.

\bibitem{H0start_10}
E. Di Valentino et al, Phys. Rev. D 101, 063502 (2020).

\bibitem{H0End}
R. C. Nunes, JCAP 05, 052 (2018).

\bibitem{Bamba:2012cp}
K. Bamba et., Astrophys. Space Sci. 342, 155 (2012). 

\bibitem{Odintsov:2017icc}
 S. D. Odintsov et al, Phys. Rev. D 96, no.4, 044022 (2017).
 
\bibitem{Yang}
W. Yang et al, arXiv:2001.02180.

\bibitem{Yang1}
W. Yang et al, JCAP 1911, 044 (2019).

\bibitem{MK8}
W. Yang et al, JCAP 1911, 044  (2019). 

\bibitem{MK12}
C. Li et al, Phys. Lett. B 80, 135141 (2020).

\bibitem{MK13}
M. Khurshudyan, R. Myrzakulov, Eur. Phys. J. C 77: 65 (2017).  

\bibitem{Mk14}
W. Yang, et al, Phys. Rev. D 99, 043543 (2019). 

\bibitem{Mk15}
E. Elizalde, M. Khurshudyan, Int. J. Mod. Phys. D 27, 1850037 (2018).

\bibitem{Mk18}
E. Sadri et al, Eur. Phys. J. C  80:393 (2020).

\bibitem{INStart}
I. Brevik et al, Int. J. Geom. Meth. Mod. Phys. 14, 1750185 (2017).

\bibitem{INStart_1}
S. Capozziello et al, Phys. Rev. D 73, 043512 (2006).

\bibitem{INStart_2}
S. Capozziello et al, Phys.Rev. D99, 023532  (2019).

\bibitem{INStart_4}
S. Nojiri, S. D. Odintsov, Phys. Rev. D 72, 023003 (2005).

\bibitem{INStart_6}
I. Brevik et al, Phys. Rev. D 86, 063007 (2012).

\bibitem{INStart_7}
I. Brevik et al, Int. J. Mod. Phys. D 26, no.14, 1730024 (2017).

\bibitem{INStart_8}
B. Mishra et al, Eur. Phys. J. C 79, 34 (2019).

\bibitem{INStart_9}
S. D. Odintsov et al, Annals Phys. 398, 238-253  (2018).

\bibitem{INStart_10}
K. Yerzhanov et al,  Mod. Phys. Lett. A 36, 2150222 (2021).

\bibitem{INStart_11}
M. Aljaf et al, arXiv:2010.05278.

\bibitem{INStart_12}
M. Khurshudyan, Symmetry 10, 577 (2018).

\bibitem{INStart_12}
S Nojiri et al, Phys. Lett. B 825, 136844 (2022).

\bibitem{INEnd}
S. D. Odintsov et al,  Phys. Rev. D 101, 044010 (2020).

\bibitem{QDE}
M. Khurshudyan, Int. Journal of Theor. Phys. 2370–2378 (2014).

\bibitem{QDE_1}
L. Arturo Urena-Lopez and Nandan Roy, Phys. Rev. D 102, 063510 (2020).

\bibitem{GP_0}
YF Cai et al, Astrophys.J. 888, 62 (2020).

\bibitem{GP_1}
E. Elizalde, M. Khurshudyan, Phys. Rev .D 99, 103533 (2019).

\bibitem{GP_2}
M. Aljaf, et al, Eur. Phys. J. C 81, 544  (2021).

\bibitem{GP_3}
E. Elizalde et al, Int. J. Mod. Phys. D 28, 1950019, (2018).

\bibitem{GP_4}
X. Rin et al, arxiv: 2203.01926.

\bibitem{GP_5}
J. L. Said et al, JCAP 06, 015  (2021).

\bibitem{Seikel}
M. Seikel, C. Clarkson and M. Smith, JCAP 06, 036 (2012).

\bibitem{Lavinia}
L. Heisenberg et al, Phys. Rev. D 98, 123502 (2018).



 
 
 
 
 \end{thebibliography}

\section*{APPENDIX}

\begin{figure}[h!]
 \begin{center}$
 \begin{array}{cccc}
\includegraphics[width=80 mm]{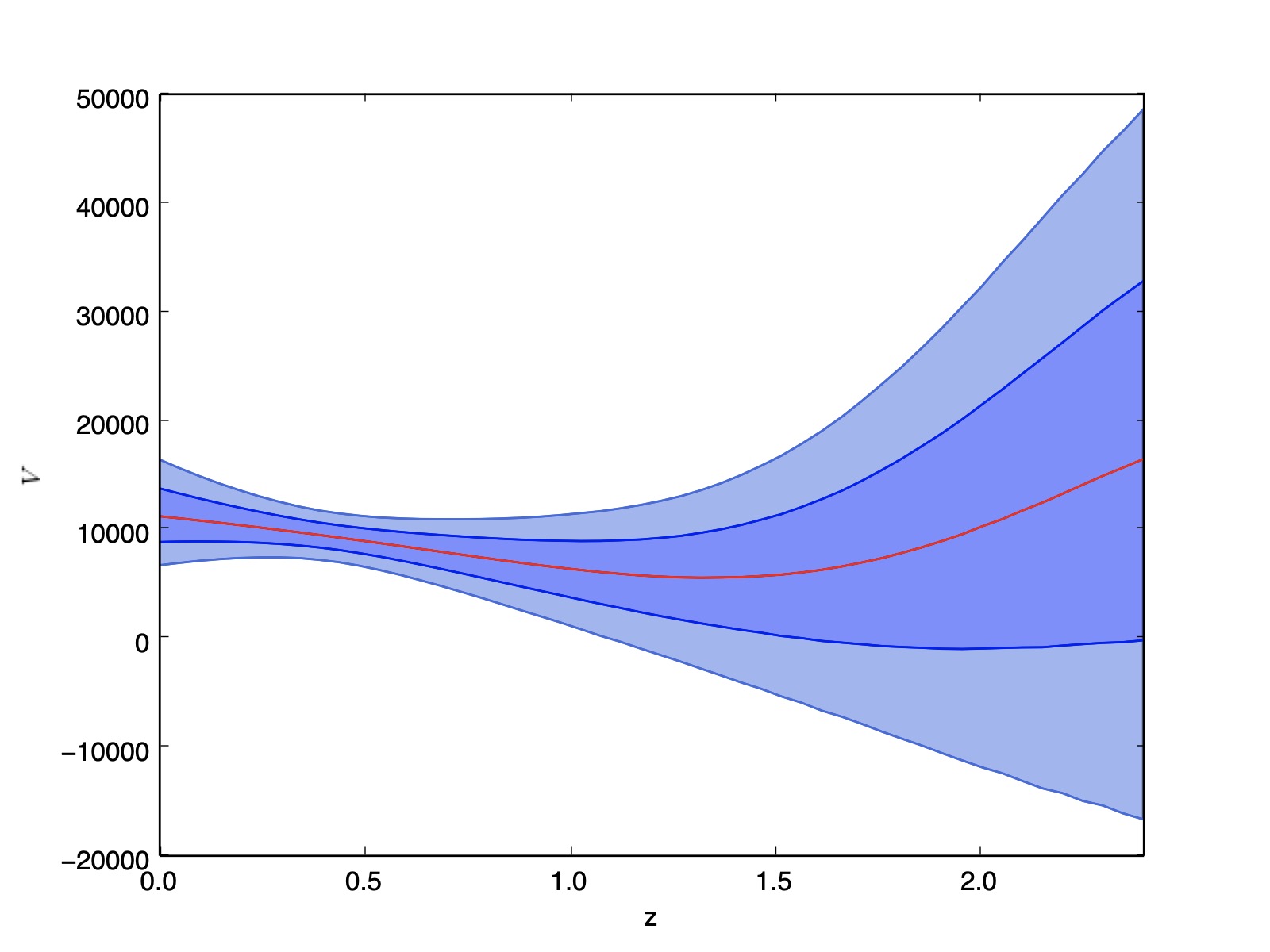}&&
\includegraphics[width=80 mm]{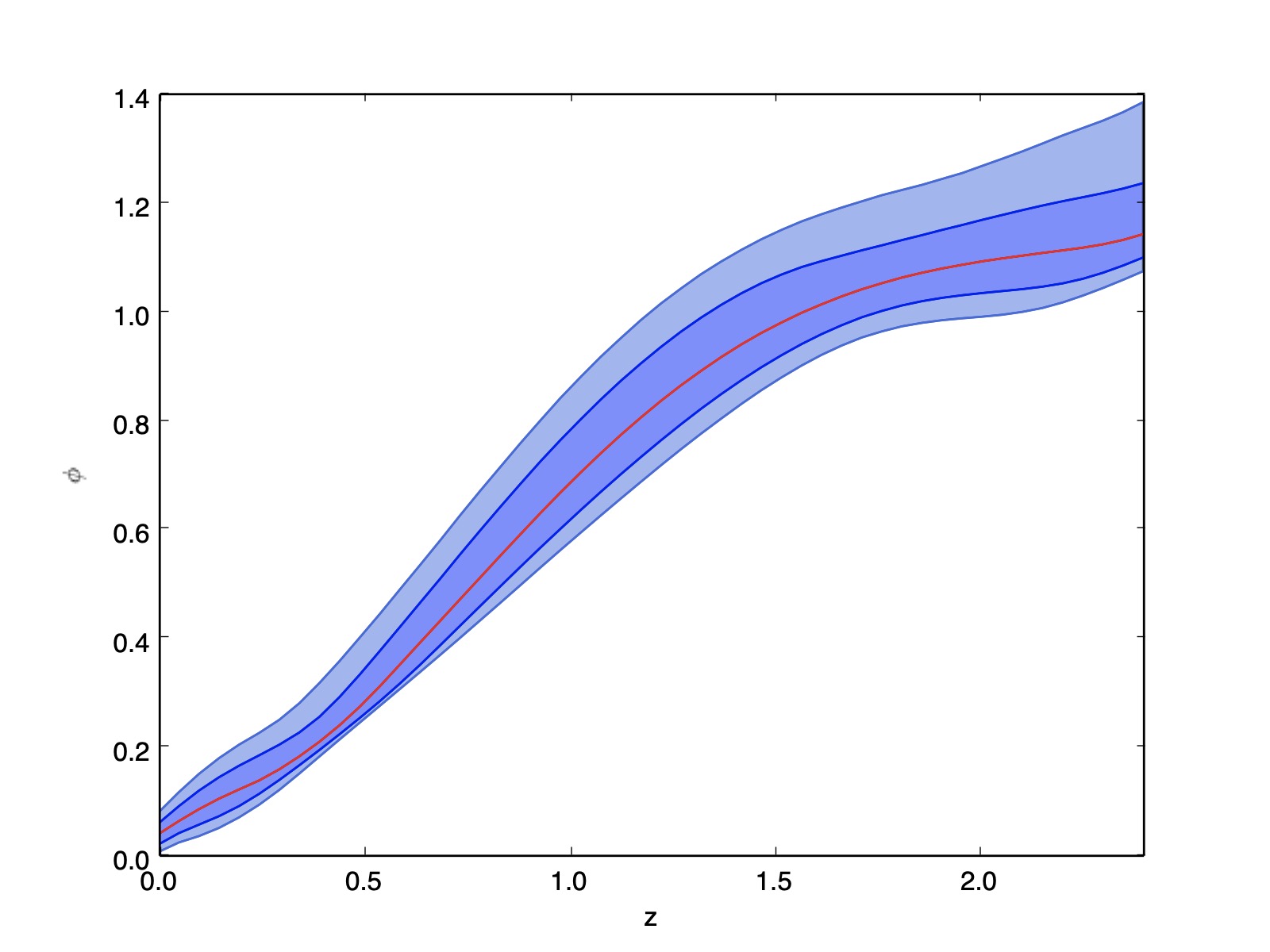}\\
\includegraphics[width=80 mm]{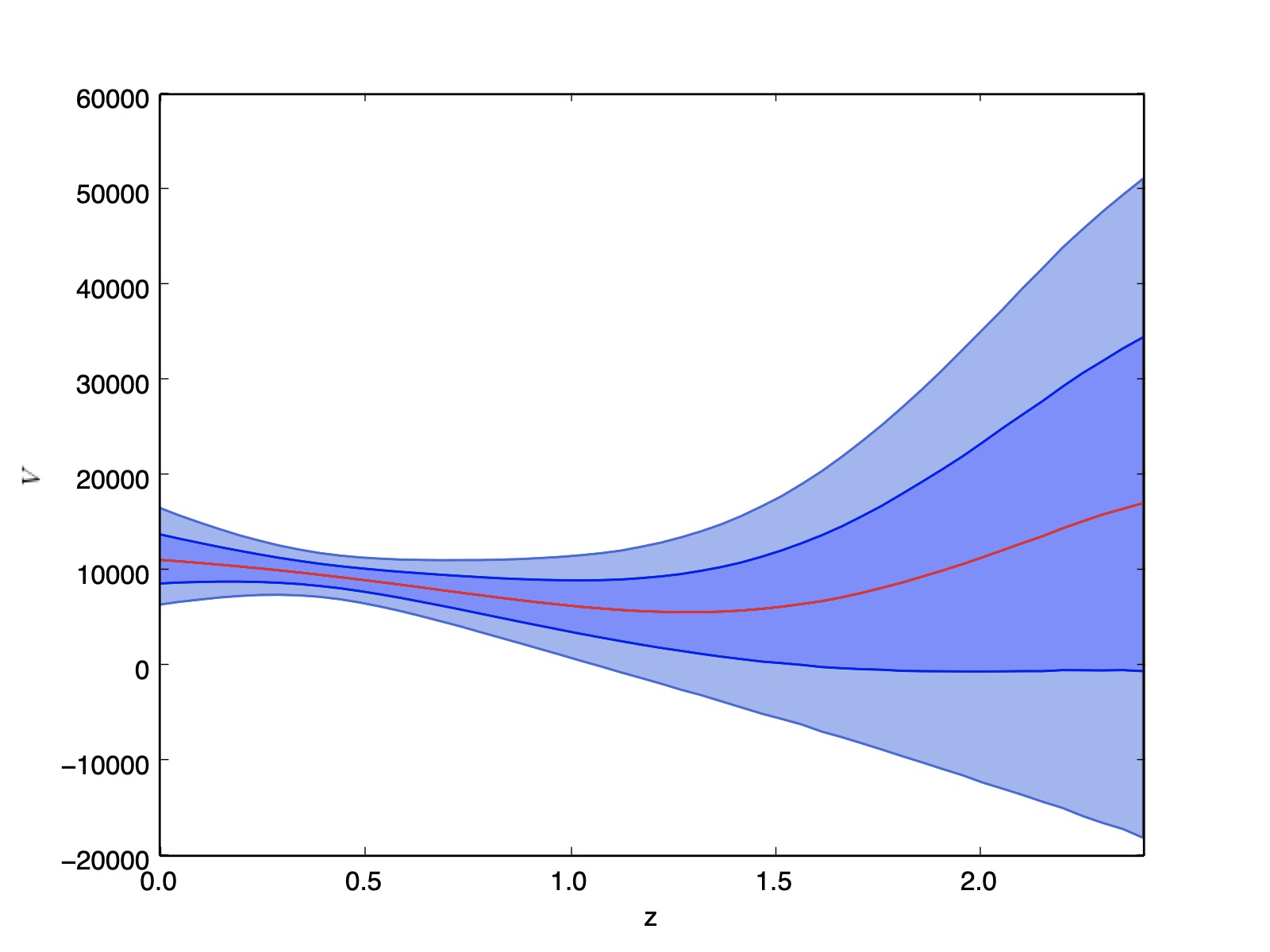}&&
\includegraphics[width=80 mm]{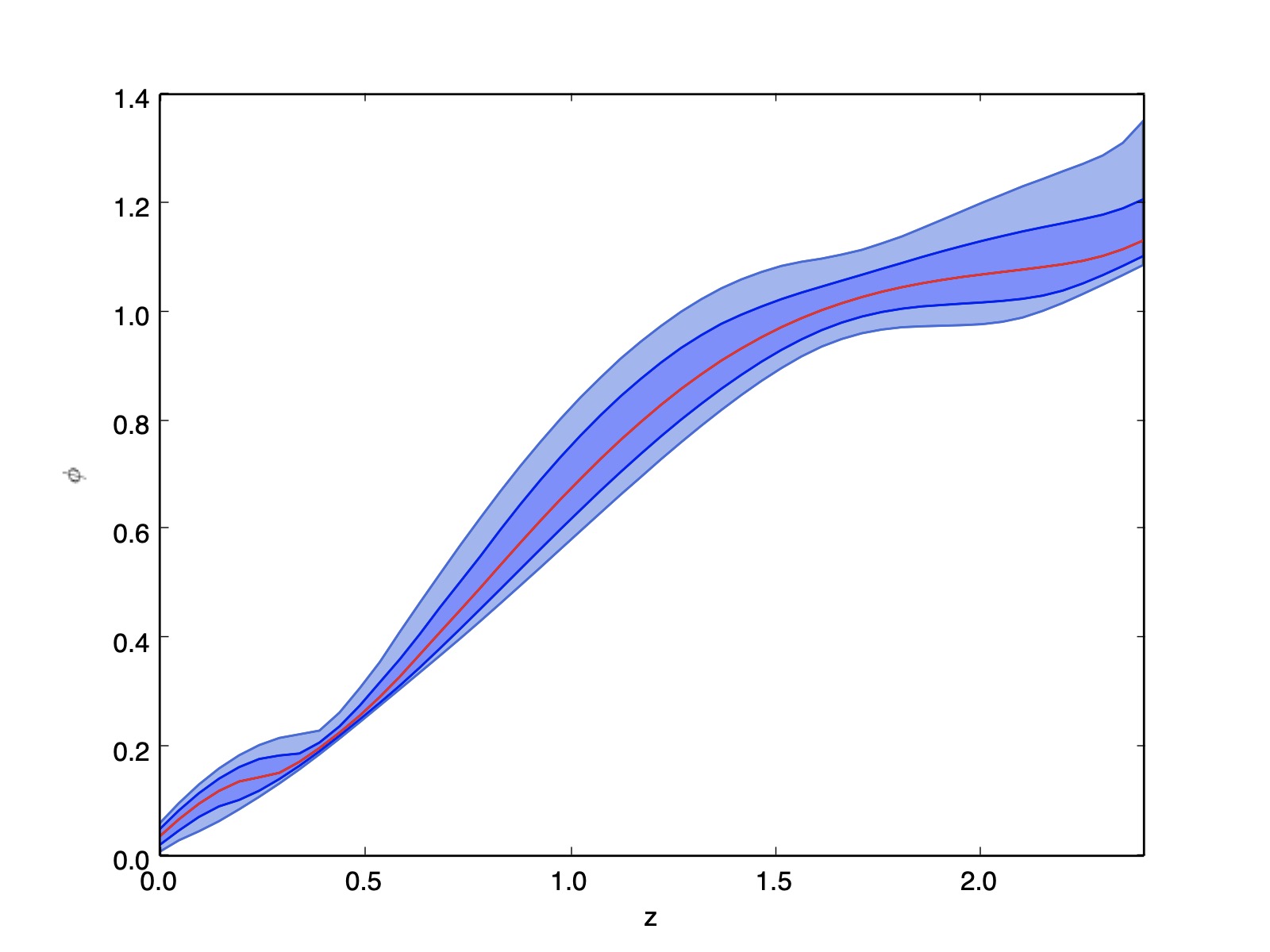}\\
 \end{array}$
 \end{center}
\caption{Reconstruction of $V(z)$, Eq. (\ref{eq:Vz}), and $\phi(z)$, Eq. (\ref{eq:phiz}), from the $H(z)$ data depicted in Table\ref{tab:Table0}. The plots of the top panel correspond to the GP reconstruction for the squared exponent kernel given by Eq. (\ref{eq:kernel1}). The plots of the bottom panel correspond to the GP reconstruction for Matern ($\nu = 9/2$) kernel given by Eq. (\ref{eq:kernel2}). The solid line is the mean of the reconstruction and the shaded blue regions are the $68\%$ and $95\%$ C.L. of the reconstruction, respectively. $H_{0} = 71.286 \pm 3.743$ km s$^{-1}$ Mpc$^{-1}$ and $H_{0} = 71.196 \pm 3.867$  km s$^{-1}$ Mpc$^{-1}$  estimates for $H_{0}$ have been obtained using GP when the kernel are given by Eq. (\ref{eq:kernel1}) and  Eq. (\ref{eq:kernel2}), respectively.}
 \label{fig:Fig2}
\end{figure}

\begin{figure}[h!]
 \begin{center}$
 \begin{array}{cccc}
\includegraphics[width=85 mm]{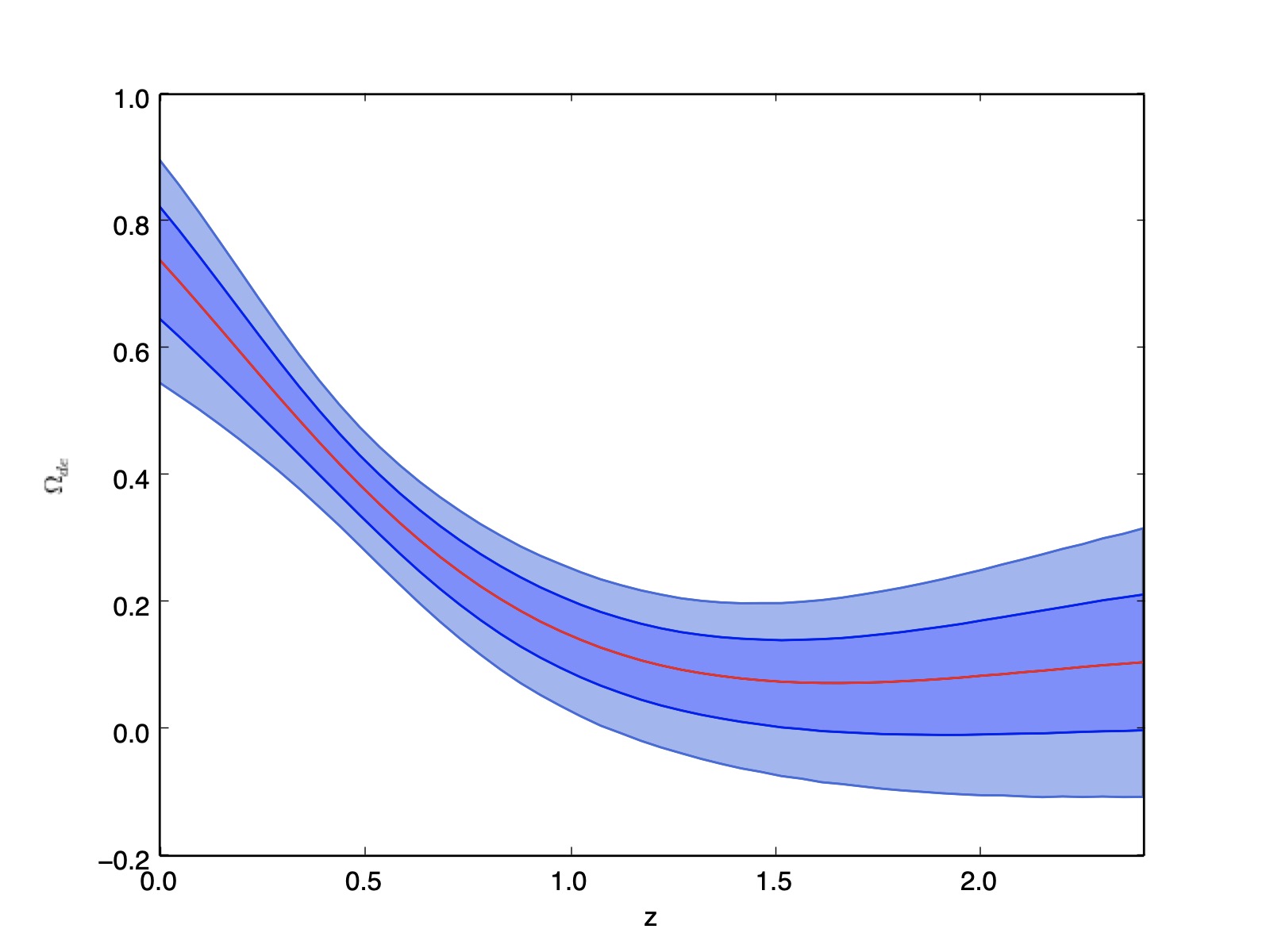}&&
\includegraphics[width=85 mm]{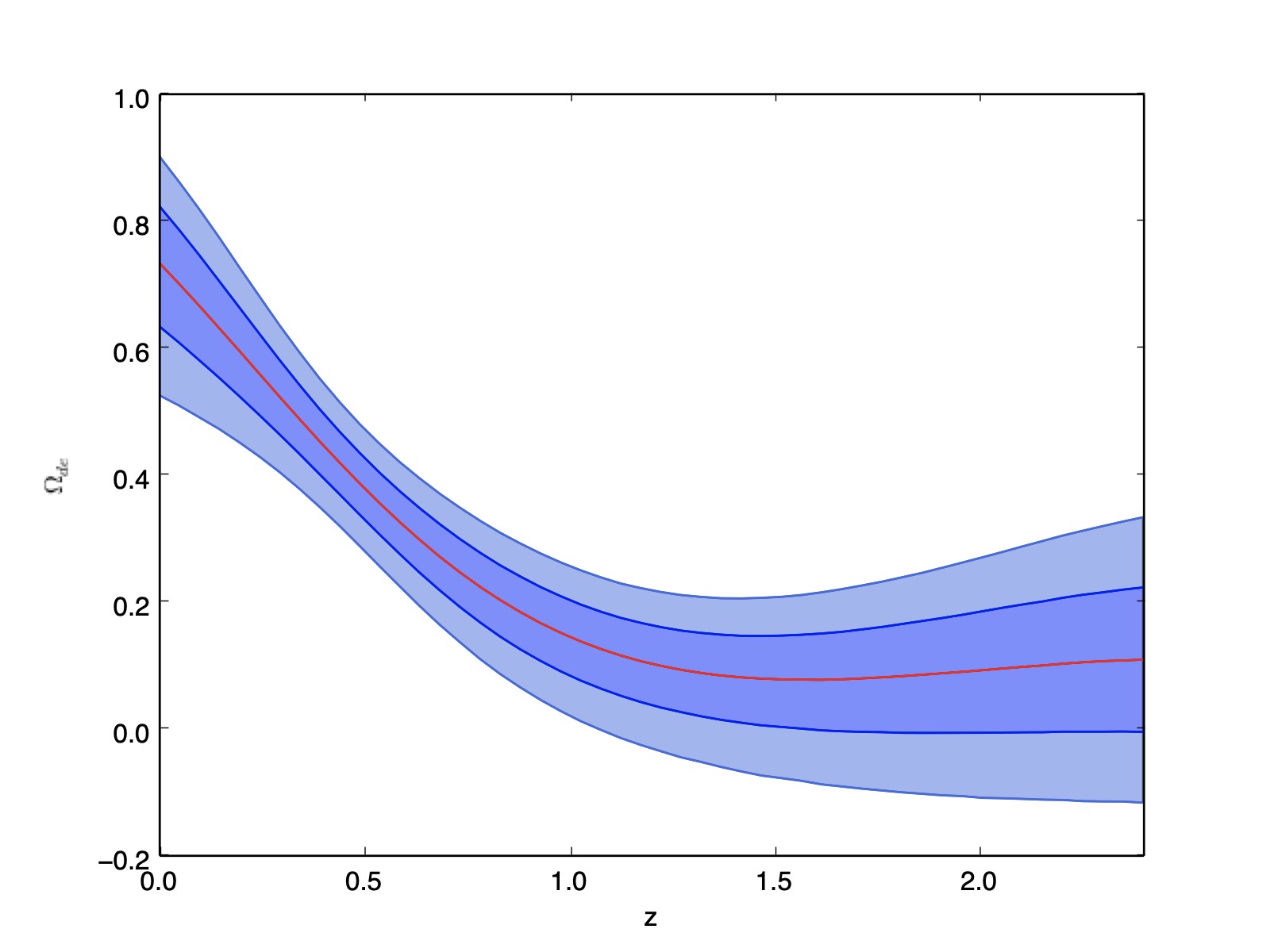}\\
 \end{array}$
 \end{center}
\caption{Reconstruction of  $\Omega_{de} = \frac{\rho_{\phi}}{3H^{2}}$ from the $H(z)$ data depicted in Table\ref{tab:Table0}. The left hand side plot corresponds to the GP reconstruction for the squared exponent kernel given by Eq. (\ref{eq:kernel1}). The right hand side plot corresponds to the GP reconstruction for the Matern ($\nu = 9/2$) kernel given by Eq. (\ref{eq:kernel2}). The solid line is the mean of the reconstruction and the shaded blue regions are the $68\%$ and $95\%$ C.L. of the reconstruction, respectively.  $H_{0} = 71.286 \pm 3.743$  km s$^{-1}$ Mpc$^{-1}$ has been estimated by GP from the data presented in Table \ref{tab:Table0} when the squared exponent kernel given by Eq. (\ref{eq:kernel1}) has been used. On the other hand, $H_{0} = 71.196 \pm 3.867$  km s$^{-1}$ Mpc$^{-1}$ has been estimated by GP from the data presented in Table \ref{tab:Table0} when the Matern ($\nu = 9/2$) kernel given by Eq. (\ref{eq:kernel2}) has been used.}
 \label{fig:FigA_2}
\end{figure}

\begin{figure}[h!]
 \begin{center}$
 \begin{array}{cccc}
\includegraphics[width=80 mm]{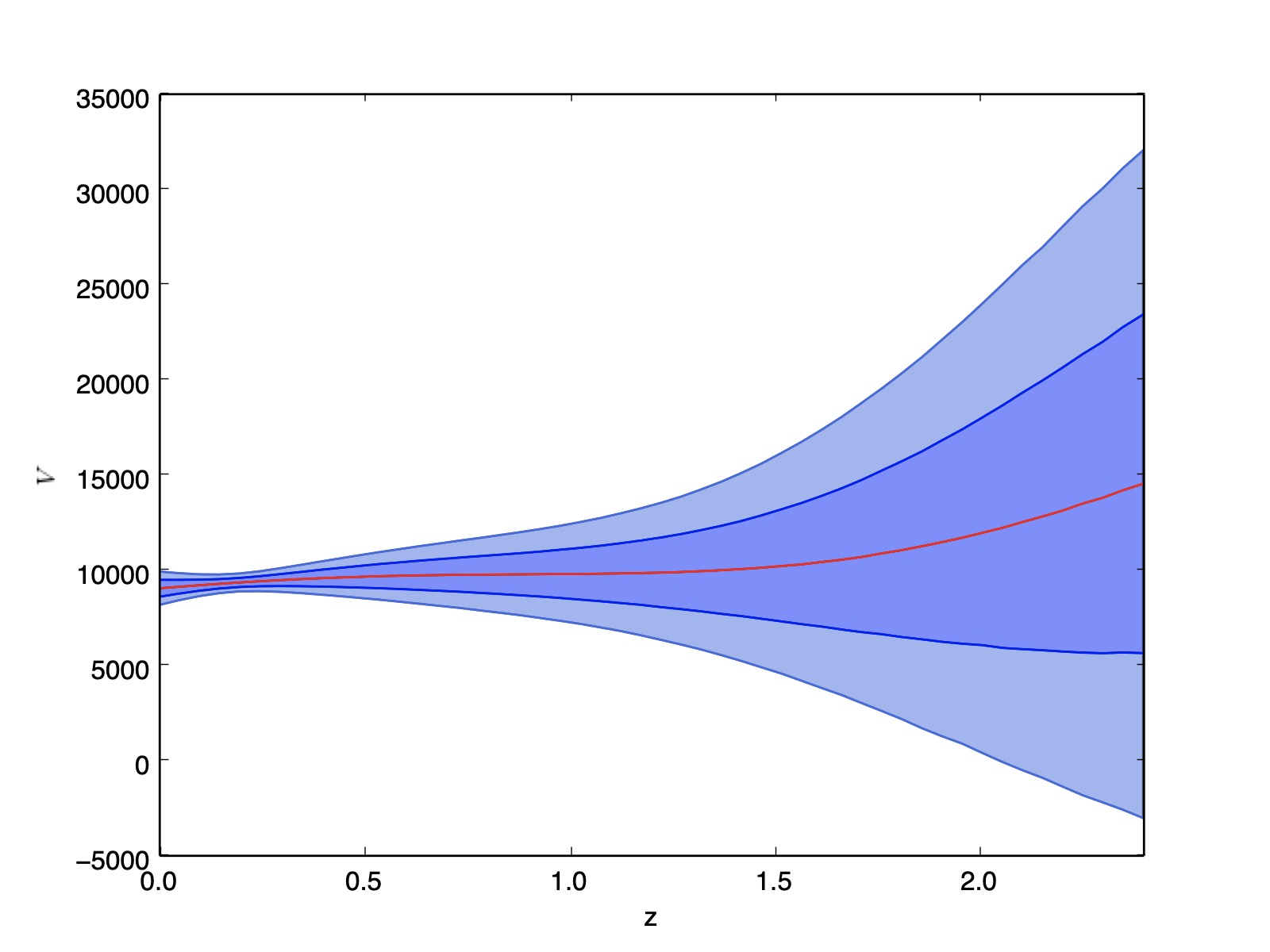}&&
\includegraphics[width=80 mm]{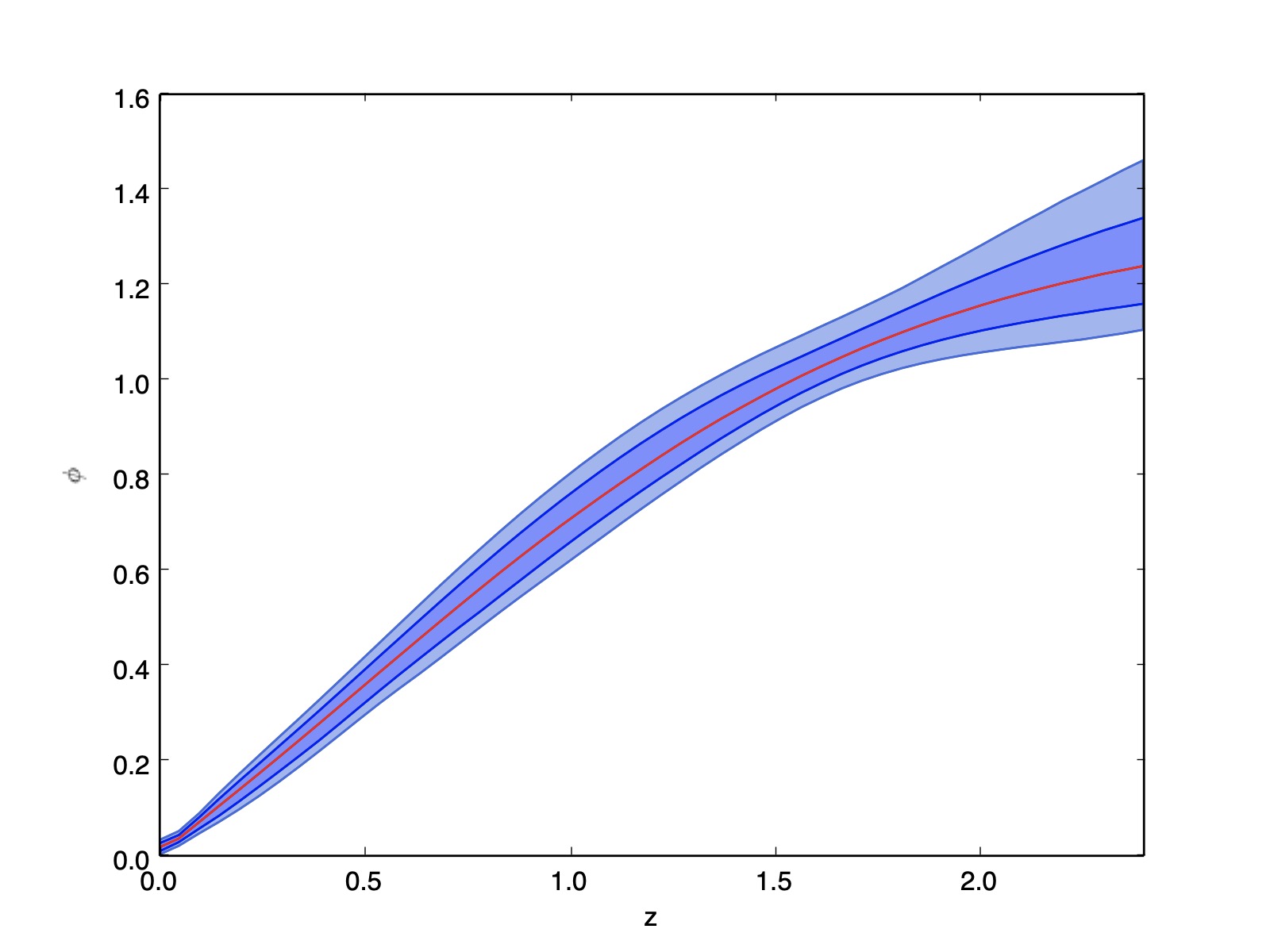}\\
\includegraphics[width=80 mm]{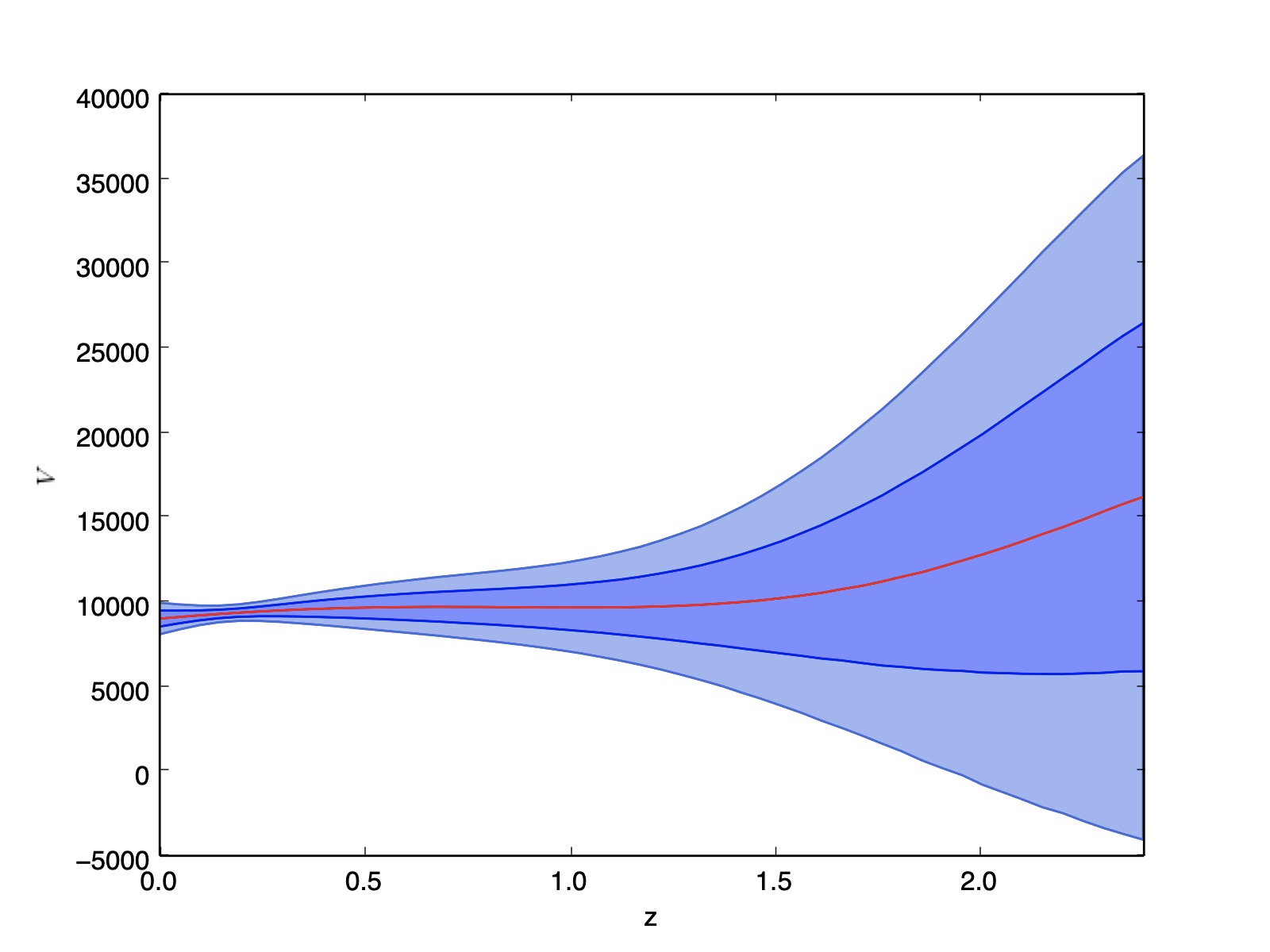}&&
\includegraphics[width=80 mm]{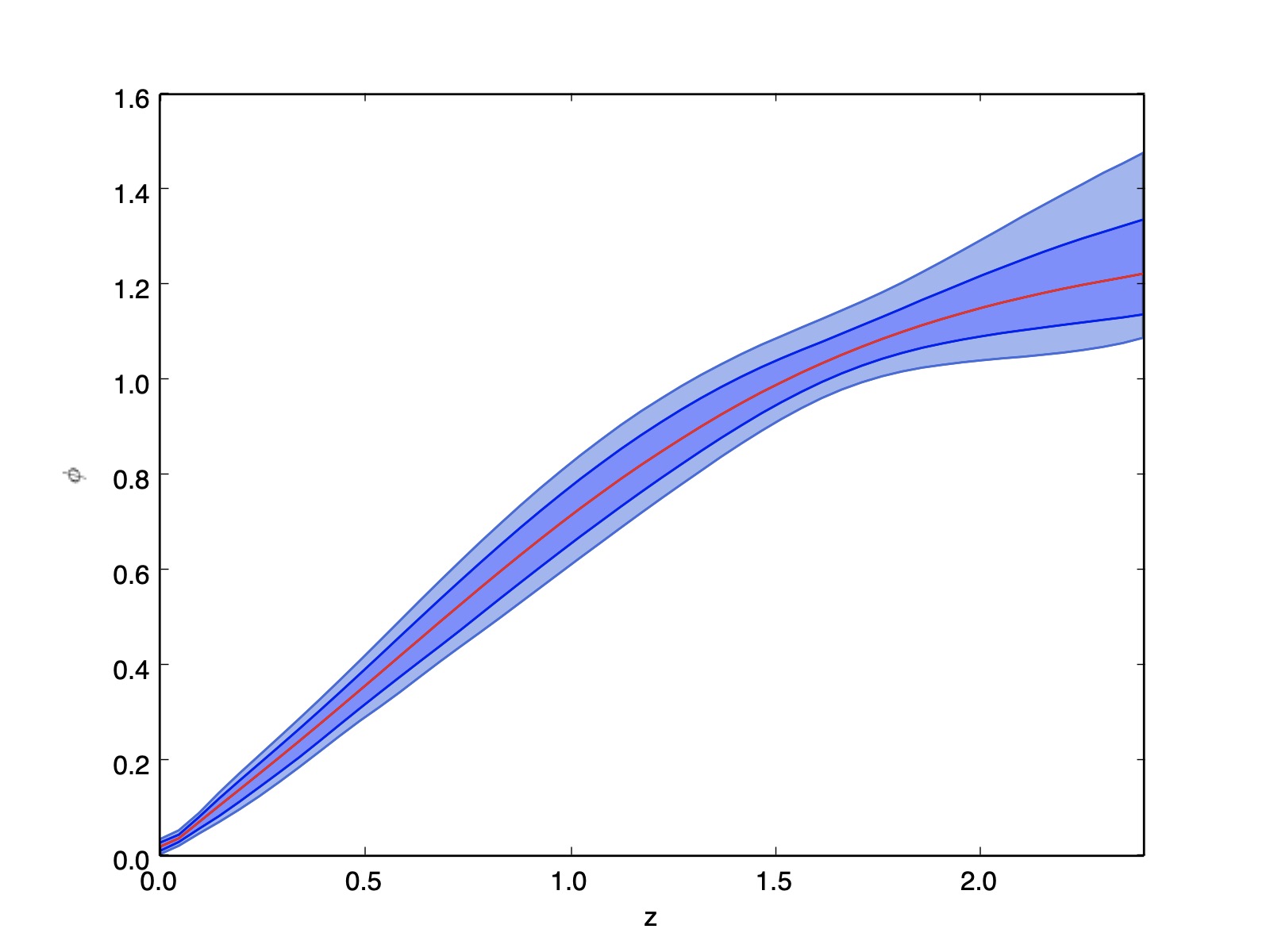}\\
 \end{array}$
 \end{center}
\caption{Reconstruction of  $V(z)$, Eq. (\ref{eq:Vz}), and $\phi(z)$, Eq. (\ref{eq:phiz}), from the $H(z)$ data depicted in Table\ref{tab:Table0} when $H_{0} = 67.40 \pm 0.5$ km s$^{-1}$ Mpc$^{-1}$. The plots of the top panel correspond to the GP reconstruction for the squared exponent kernel, Eq. (\ref{eq:kernel1}). The plots of the bottom panel correspond to the GP reconstruction for the kernel given by Eq. (\ref{eq:kernel1}).  The solid line is the mean of the reconstruction and the shaded blue regions are the $68\%$ and $95\%$ C.L. of the reconstruction, respectively.}
 \label{fig:Fig3}
\end{figure}

\begin{figure}[h!]
 \begin{center}$
 \begin{array}{cccc}
\includegraphics[width=80 mm]{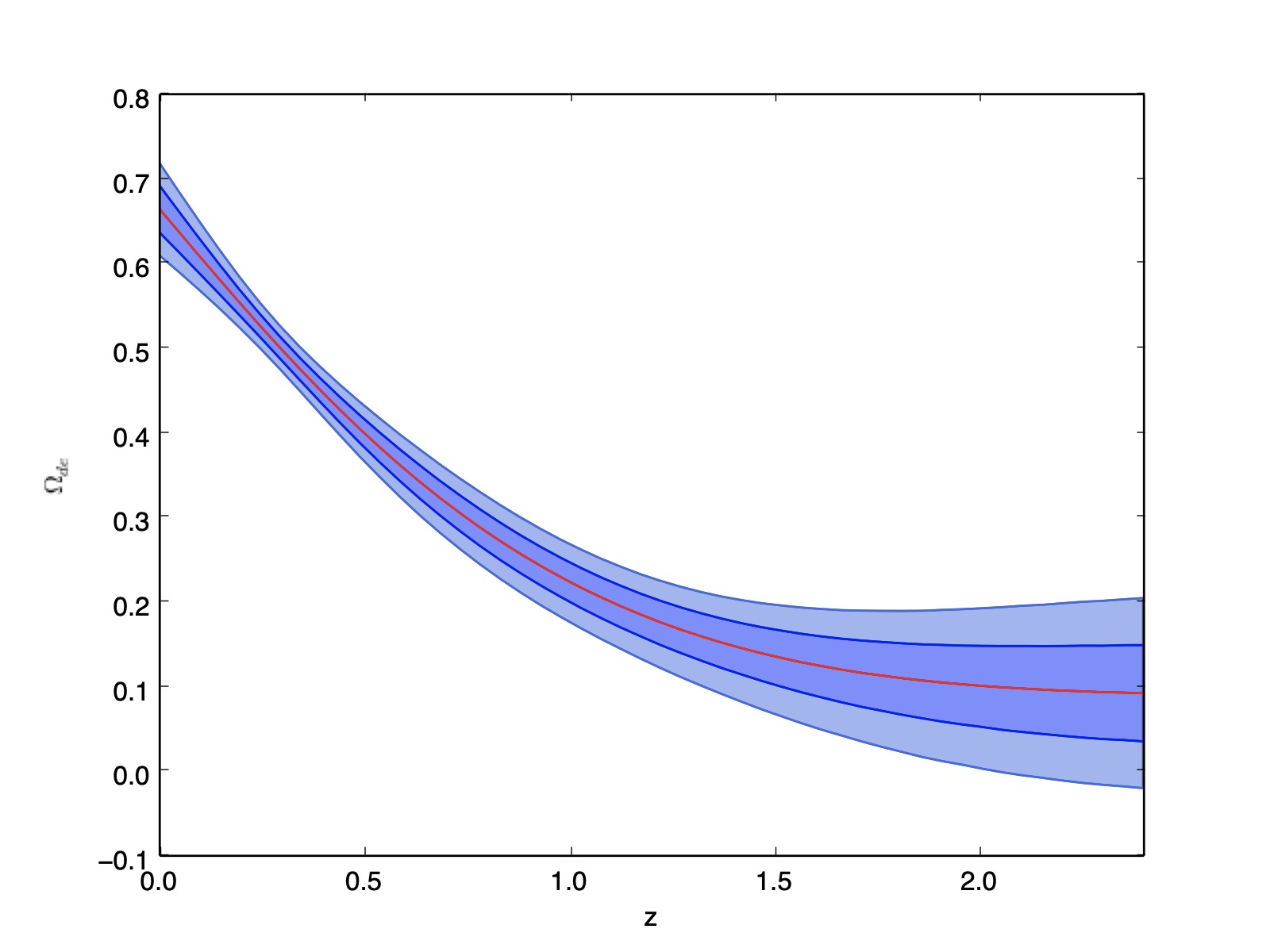}&&
\includegraphics[width=80 mm]{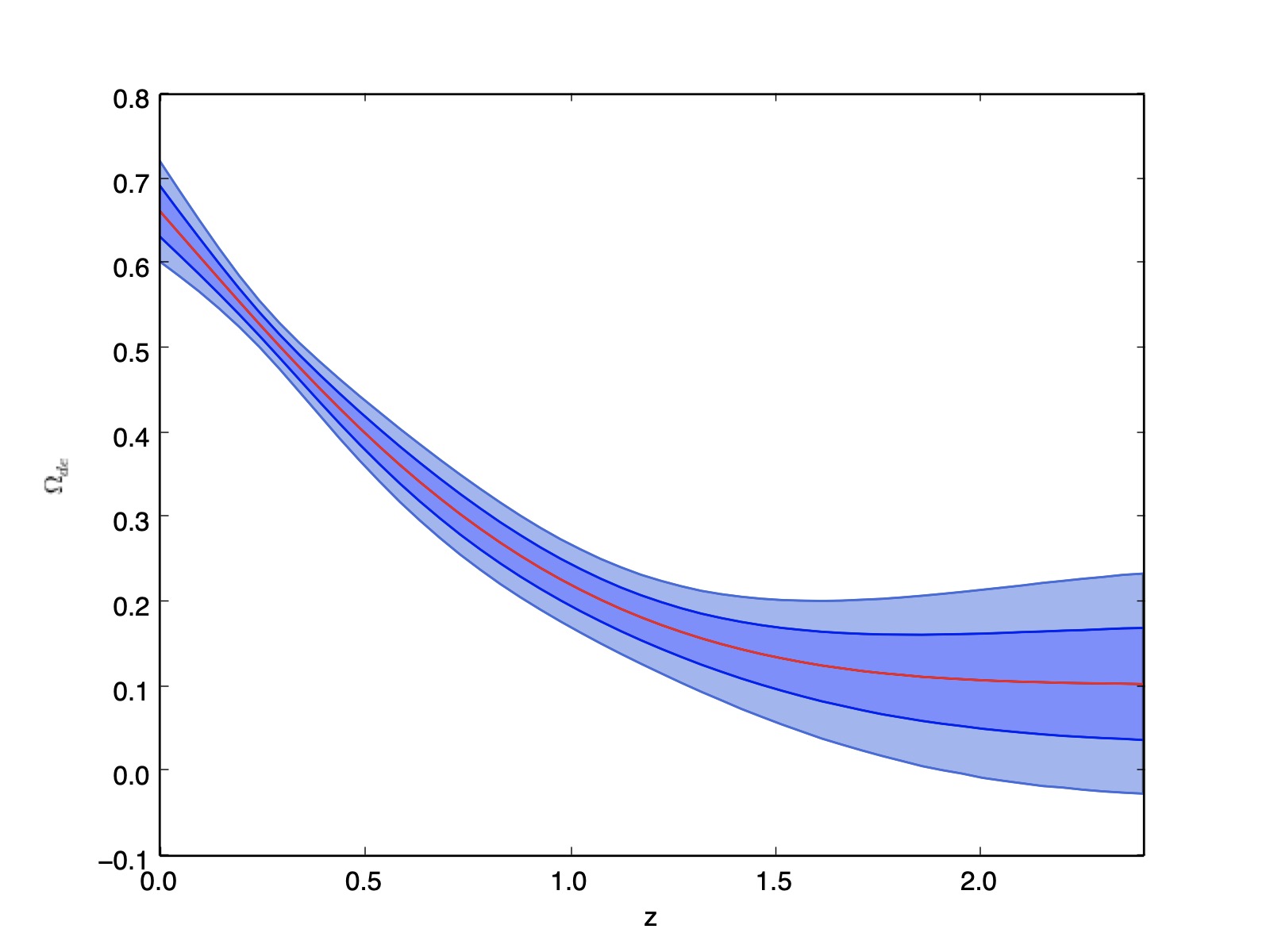}\\
 \end{array}$
 \end{center}
\caption{Reconstruction of  $\Omega_{de} = \frac{\rho_{\phi}}{3H^{2}}$ from the $H(z)$ data depicted in Table\ref{tab:Table0} when $H_{0} = 67.40 \pm 0.5$ km s$^{-1}$ Mpc$^{-1}$. The left hand side plot corresponds to the GP reconstruction for the squared exponent kernel given by Eq. (\ref{eq:kernel1}). The right hand side plot correspond to the GP reconstruction for the Matern ($\nu = 9/2$) kernel given by Eq. (\ref{eq:kernel2}). The solid line is the mean of the reconstruction and the shaded blue regions are the $68\%$ and $95\%$ C.L. of the reconstruction, respectively.}
 \label{fig:FigA_3}
\end{figure}

\begin{figure}[h!]
 \begin{center}$
 \begin{array}{cccc}
\includegraphics[width=80 mm]{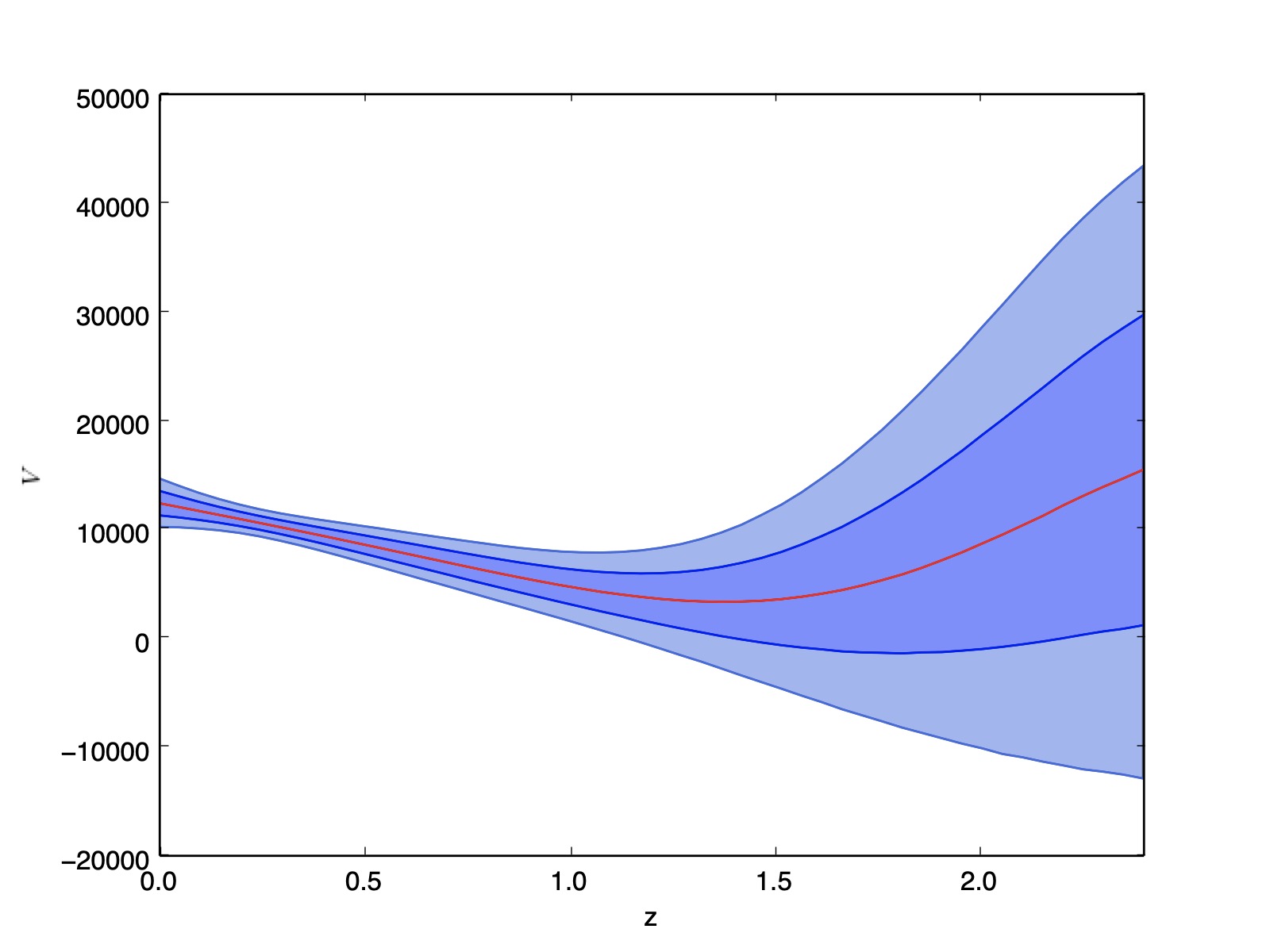}&&
\includegraphics[width=80 mm]{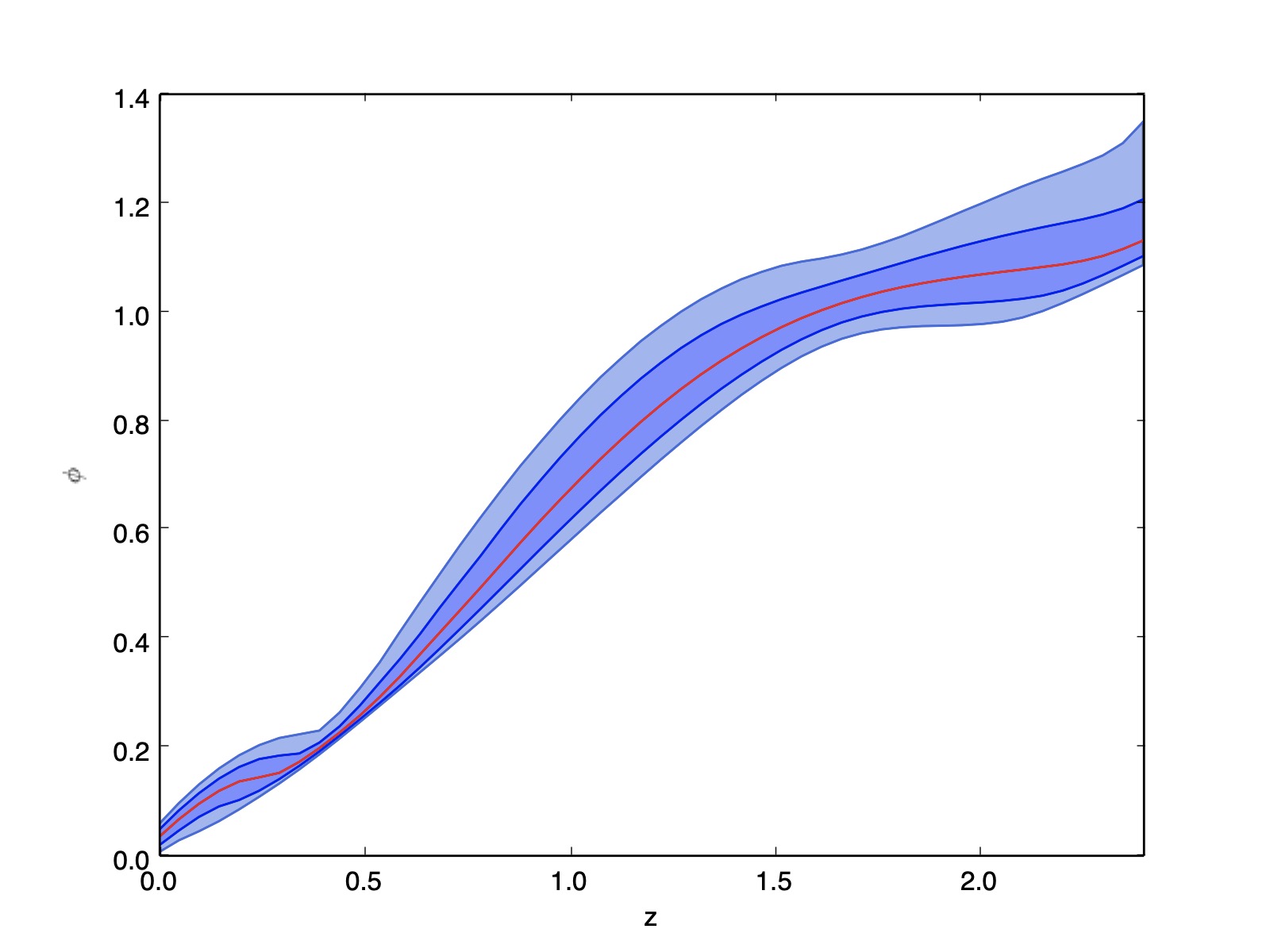}\\ 
\includegraphics[width=80 mm]{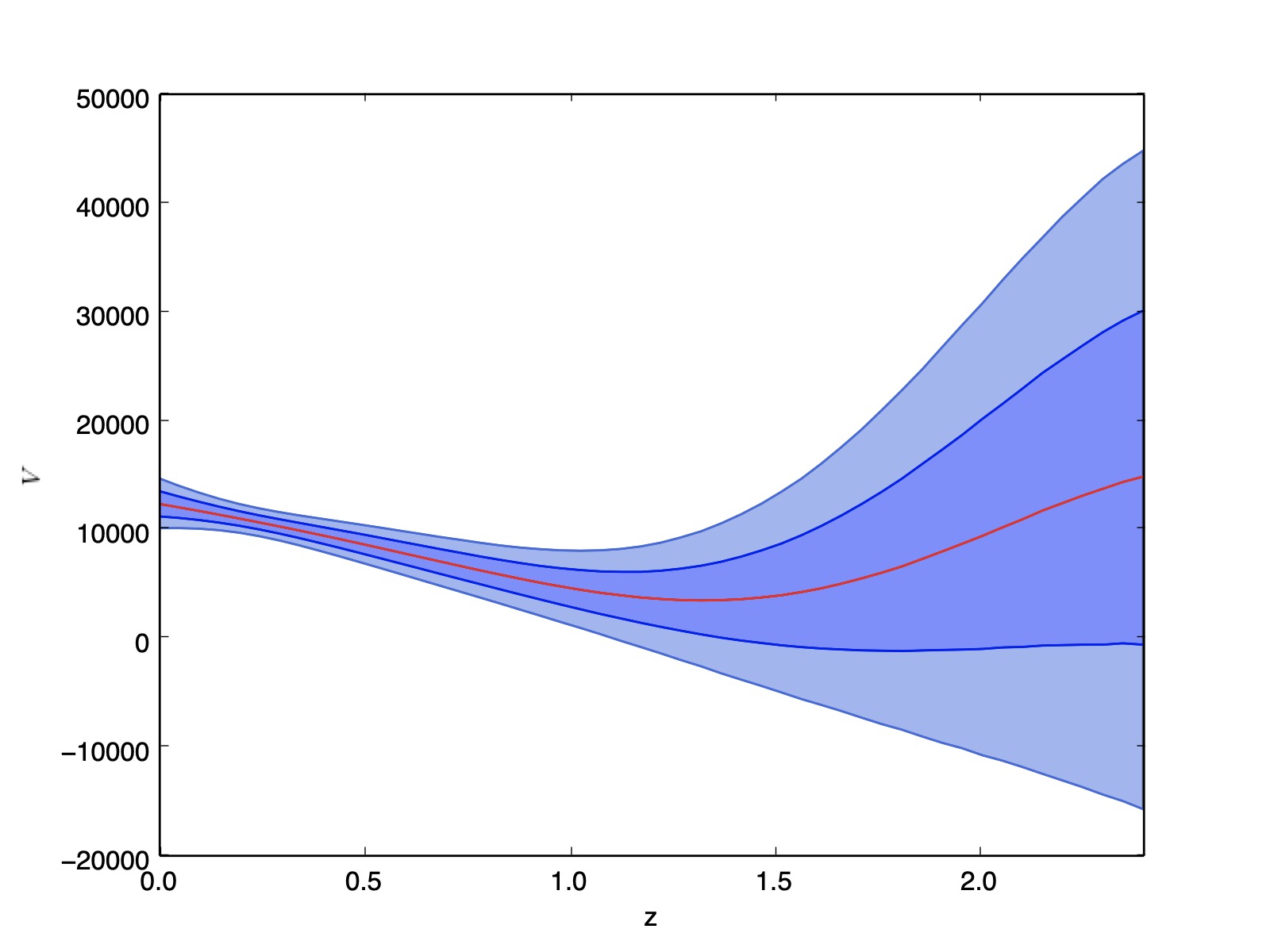}&&
\includegraphics[width=80 mm]{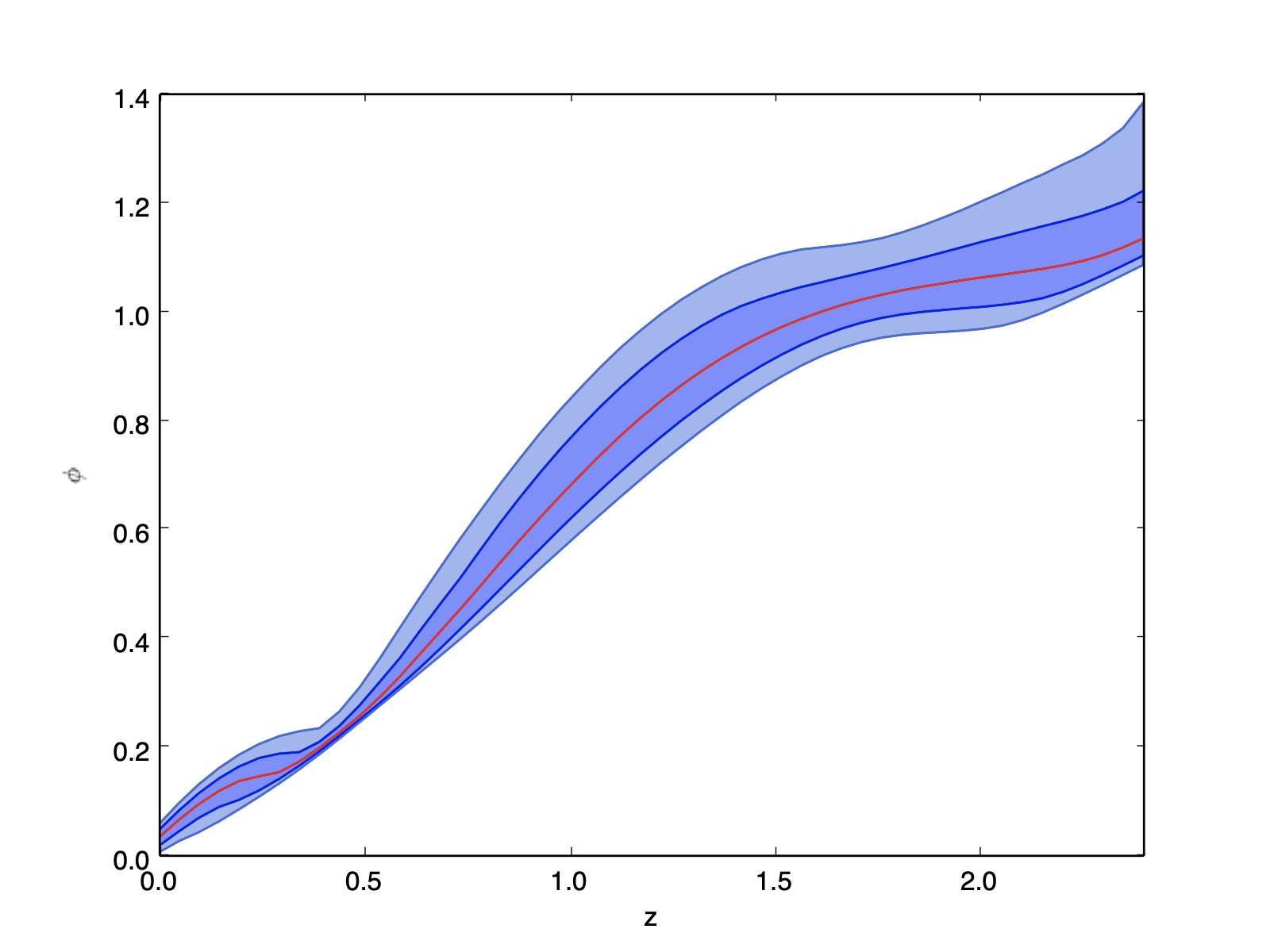}\\
 \end{array}$
 \end{center}
\caption{Reconstruction of  $V(z)$, Eq. (\ref{eq:Vz}), and $\phi(z)$, Eq. (\ref{eq:phiz}), from the $H(z)$ data depicted in Table\ref{tab:Table0} when $H_{0} = 73.52 \pm 1.62$ km s$^{-1}$ Mpc$^{-1}$. The plots of the top panel correspond to the GP reconstruction for the squared exponent kernel, Eq. (\ref{eq:kernel1}). The plots of the bottom panel correspond to the GP reconstruction for the kernel given by Eq. (\ref{eq:kernel1}).  The solid line is the mean of the reconstruction and the shaded blue regions are the $68\%$ and $95\%$ C.L. of the reconstruction, respectively.}
 \label{fig:Fig4}
\end{figure}

\begin{figure}[h!]
 \begin{center}$
 \begin{array}{cccc}
\includegraphics[width=80 mm]{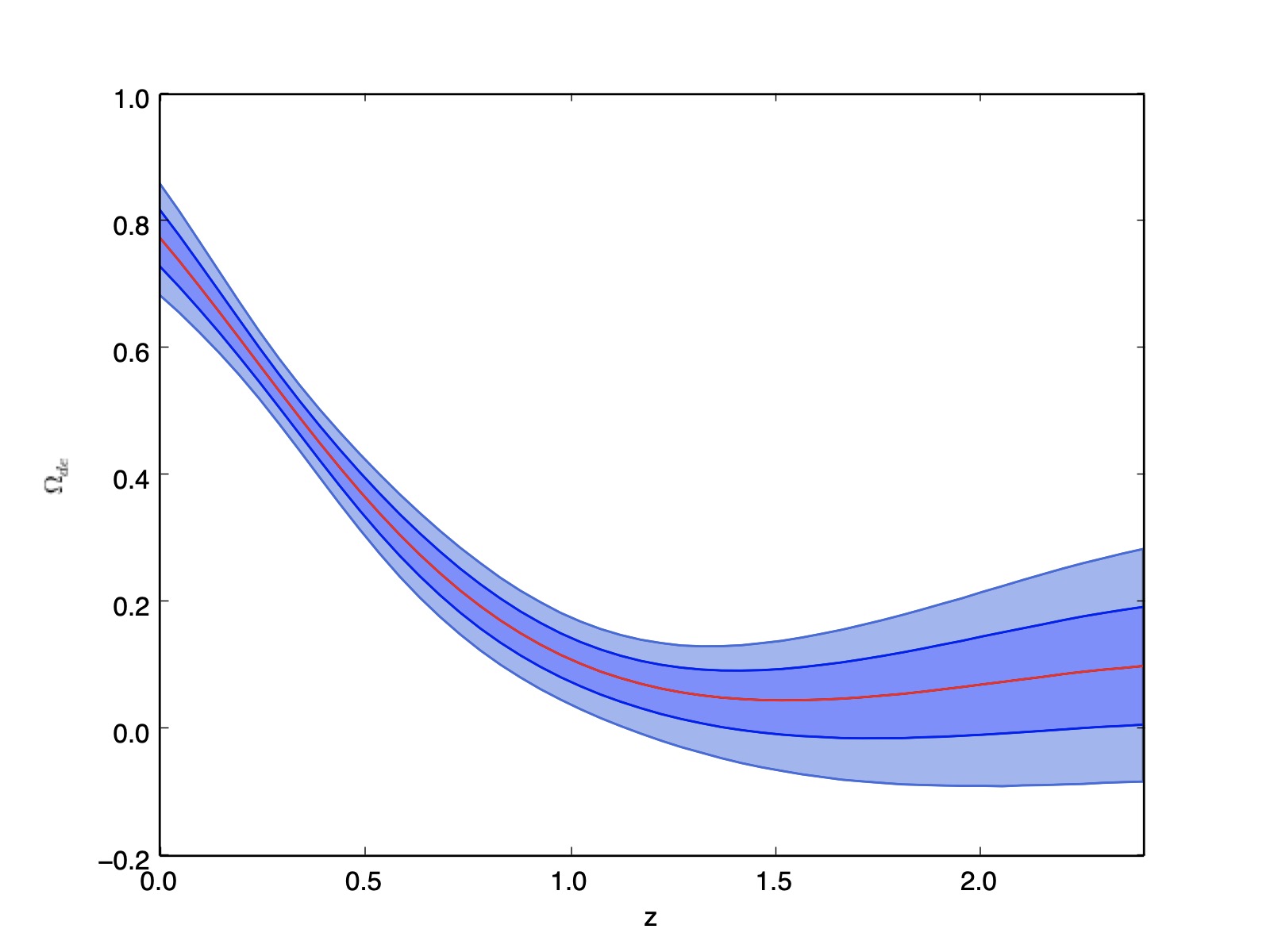}&&
\includegraphics[width=80 mm]{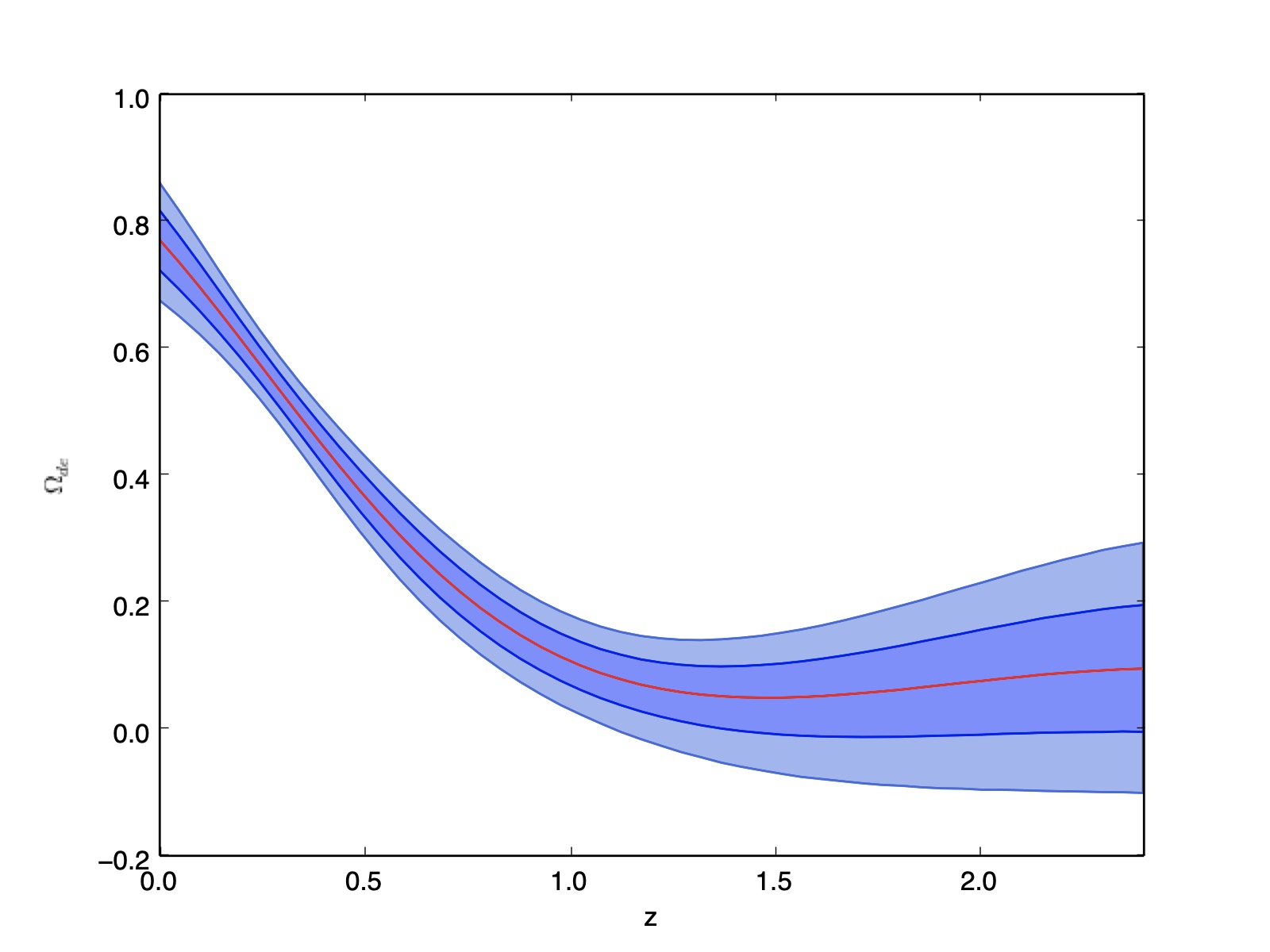}\\
 \end{array}$
 \end{center}
\caption{Reconstruction of  $\Omega_{de} = \frac{\rho_{\phi}}{3H^{2}}$ from the $H(z)$ data depicted in Table\ref{tab:Table0}, when $H_{0} = 73.52 \pm 1.62$ km s$^{-1}$ Mpc$^{-1}$. The left hand side plot corresponds to the GP reconstruction for the squared exponent kernel given by Eq. (\ref{eq:kernel1}). The right hand side plot corresponds to the GP reconstruction for the Matern ($\nu = 9/2$) kernel given by Eq.(\ref{eq:kernel2}). The solid line is the mean of the reconstruction and the shaded blue regions are the $68\%$ and $95\%$ C.L. of the reconstruction, respectively.}
 \label{fig:FigA_4}
\end{figure}

We include here several additional figures, Figs. (\ref{fig:Fig2}) - (\ref{fig:FigA_4}), to allow the reader to estimate the quality of the reconstructions presented in the paper.

\end{document}